\pdfoutput=1
\documentclass[aps,prd,nofootinbib,showkeys,superscriptaddress,twocolumn,floatfix]{revtex4-1}

\usepackage{amssymb}
\usepackage{amsmath}
\usepackage[colorlinks=true,linkcolor=blue,citecolor=blue, urlcolor=blue]{hyperref}   
\usepackage{tikz}
\usepackage{graphicx}

\newcommand{\kompost}{K{\o}MP{\o}ST}

\definecolor{uibred}{RGB}{167, 38, 47}

\newcommand{\xt}{\mathbf{x}}

\newcommand{\kt}{\mathbf{k}}

\newcommand{\pt}{\mathbf{p}}

\newcommand{\thetaP}{\theta_{\mathbf{p}}}

\def\st{\begin{equation}}
\def\stp{\end{equation}}

\usepackage{tikz}

\begin{document}

\title{Hydrodynamization and non-equilibrium Green's functions in kinetic theory}

\author{Syo Kamata}
\email[]{skamata11phys@gmail.com}
\affiliation{College of Physics and Communication Electronics, Jiangxi Normal University, Nanchang 330022, China
}
\affiliation{Department
of Physics, North Carolina State University, Raleigh, NC 27695, USA}

\author{Mauricio Martinez}
\email[]{mmarti11@ncsu.edu}
\affiliation{Department
of Physics, North Carolina State University, Raleigh, NC 27695, USA}

\author{Philip Plaschke}
\email[]{p.plaschke@uni-bielefeld.de}
\affiliation{Fakult\"{a}t f\"{u}r Physik, Universit\"{a}t Bielefeld, D-33615 Bielefeld, Germany}

\author{Stephan Ochsenfeld}
\email[]{s.ochsenfeld@uni-bielefeld.de}
\affiliation{Fakult\"{a}t f\"{u}r Physik, Universit\"{a}t Bielefeld, D-33615 Bielefeld, Germany}

\author{S\"{o}ren Schlichting}
\email[]{sschlichting@physik.uni-bielefeld.de}
\affiliation{Fakult\"{a}t f\"{u}r Physik, Universit\"{a}t Bielefeld, D-33615 Bielefeld, Germany}

\date{\today}

\begin{abstract}
Non-equilibrium Green's functions provide an efficient way to describe the evolution of the energy-momentum tensor during the early time pre-equilibrium stage of high-energy heavy ion collisions. Besides their practical relevance they also provide a meaningful way to address the question when and to what extent a hydrodynamic description of the system becomes applicable. Within the kinetic theory framework we derive a new method to calculate time dependent non-equilibrium Green's functions describing the evolution of energy and momentum perturbations on top of an evolving far-from-equilibrium background. We discuss the approach towards viscous hydrodynamics along with the emergence of various scaling phenomena for conformal systems. By comparing our results obtained in the relaxation time approximation to previous calculations in Yang-Mills kinetic theory, we further address the question which macroscopic features of the energy momentum tensor are sensitive to the underlying microscopic dynamics. 
\end{abstract}

\maketitle

\section{Introduction}
Ultrarelativistic heavy ion experiments carried out at the Relativistic Heavy Ion Collider (RHIC) and the Large Hadron Collider (LHC) have produced a new state of matter where quarks and gluons are liberated from the incoming nuclei~\cite{Adams:2005dq,Adcox:2004mh,Back:2004je,Arsene:2004fa} Since the lifetime of this Quark-Gluon Plasma (QGP) is very short, its properties are reconstructed by analyzing a large set of hadronic observables. The phenomenological studies of the collected wealth of experimental data have shown that hydrodynamical models provide robust tools to explain the dynamics of the QGP~\cite{Heinz:2013th,Luzum:2013yya,Gale:2013da,deSouza:2015ena,Teaney:2009qa}. As a result, a new paradigm in high energy nuclear physics has emerged where hydrodynamics plays a central role. 

One of the most consistent findings in the hydrodynamical modeling of heavy ion collisions is a small value of the shear viscosity over entropy ratio $\eta/s\sim 1/(4\pi)$, extracted from large number of observables measured in ultrarelativistic heavy ion collisions over a different range of energies at both RHIC and LHC. Currently, the standard approach to describe high-energy heavy ion collisions is based on multistage models where the space-time dynamics of the QGP is described  using relativistic viscous hydrodynamics. Subsequently, the energy density of the QGP is converted into a hadron gas which eventually freezes out, yielding the final state hadronic observables that can compared to experimental measurements. One key ingredient in this phenomenological approach are the initial conditions, characterizing the initial distributions of the energy density and flow velocities, which are then propagated via hydrodynamics. Clearly, the calculation of these quantities from  first principles QCD represents an enormous challenge, which despite important developments during the last years~\cite{Kharzeev:2001yq,Kharzeev:2001yq,Moreland:2014oya,Alver:2008aq,Kurkela:2014tea,vanderSchee:2013pia,Schenke:2012wb,Nagle:2018ybc,Martinez:2019jbu,Martinez:2019rlp} has not been answered completely. 

One important feature shared by all initial state models is related to the far-from-equilibrium nature of the QCD matter, produced immediately after the collision of heavy nuclei on a time scale $\tau_0 \ll 1 \rm{fm}/c$~\cite{Schlichting:2019abc}. Since at these early times the QCD matter is also subject to a rapid longitudinal expansion, viscous hydrodynamics which is an effective theory for the long-time and long wave-length behavior close to equilibrium is not necessarily applicable at such early times. Although in recent years the existence of a far-from-equilibrium fluid dynamical theory has been advocated~\cite{Romatschke:2017vte,Florkowski:2017olj,Romatschke:2017ejr,Kurkela:2015qoa,Critelli:2017euk,Denicol:2014xca,Florkowski:2013lza,Florkowski:2013lya,Denicol:2014tha,Chesler:2009cy,Heller:2011ju,vanderSchee:2012qj,Chesler:2016ceu,Martinez:2010sc,Florkowski:2010cf,Blaizot:2019scw,Blaizot:2017ucy,Blaizot:2017lht,Kurkela:2019set,Heller:2016rtz,Heller:2015dha,Jaiswal:2019cju,Casalderrey-Solana:2017zyh}, its formulation remains to be completed and thus, in practice hydrodynamic simulations of heavy-ion collisions are usually initialized after a certain period time $\tau=\tau_{\rm hydro} \sim 1 \rm{fm}/c$, where the long. expansion is less rapid and the QGP has evolved towards local thermal equilibrium. 

The question how to match the non-equilibratium initial state at $\tau_0$ to the initial state for hydrodynamics at $\tau_{\rm hydro}$ was the subject of a series of recent papers~\cite{Keegan:2016cpi,Kurkela:2018vqr,Kurkela:2018wud,Giacalone:2019ldn}. Based on an underlying microscopic description in QCD kinetic theory, the initial conditions for the energy-momentum tensor $T^{\mu\nu}(\tau_{\rm hydro})$ at the beginning of the hydrodynamic phase are hereby obtained from the initial energy-momentum tensor $T^{\mu\nu}(\tau_0)$,  by determining the evolution of macroscopic quantities, based on \emph{non-equilibrium Green's functions of the energy momentum tensor}~\cite{Kurkela:2018vqr,Kurkela:2018wud}. This new framework, dubbed as \kompost~\cite{Kurkela:2018vqr,Kurkela:2018wud}, has proven to be a powerful tool to describe the pre-equilibrium evolution of heavy-ion collisions on an event-by-event basis~\cite{Gale:2020xlg,Schenke:2020uqq}. While the detailed phenomenological consequences of the pre-equilibrium stage remain to be explored, it has been demonstrated that the subsequent hydrodynamic evolution becomes independent of the hydrodynamic initialization time $\tau_{hydro}$ as long as the latter is chosen to be in the regime where both kinetic theory and hydrodynamic descriptions overlap. 

While~\cite{Kurkela:2018vqr,Kurkela:2018wud} presented a calculation of non-equilibrium Green's functions in QCD kinetic theory, it is also interesting to explore to what extent the microscopic details affect the evolution of the macroscopic quantities far-from-equilibrium. In this work we therefore present a new method to calculate non-equilibrium Green's functions of the energy momentum tensor. We generalize the method of moments approach~\cite{DeGroot:1980dk,Grad} by incorporating the response of a far-from-equilibrium expanding plasma against linear perturbations, and analyze their non-equilibrium in the relaxation time approximation. Instead of solving linear kinetic theory for perturbations of the phase-space distribution function we analyze the equations of motion of the corresponding linearized moments. Eventually, the Green's functions of the energy momentum tensor are reconstructed from the linearized moments. 

This paper is organized as follows: In Sect.~\ref{sec:boltz} we introduce the relevant aspects of the Boltzmann equation. We explain the main aspects of the emergent attracting behavior of the non-equilibrated Bjorken flow background in Sect.~\ref{sec:back}. In Sect.~\ref{sec:EMpert} we present the formalism to describe the space-time evolution of the perturbations. Using this formalism we proceed in Sect.~\ref{sec:GreensFunctions} to calculate the Green's functions of the energy-momentum tensor. Conclusions and outlook are discussed in Sect.~\ref{sec:concl}. Some of technical aspects of our work are briefly discussed in Appendix~\ref{sec:SphericalHarmonicIdentities}.

\section{Boltzmann equation within the relaxation time approximation}
\label{sec:boltz}

Starting point of our analysis is the Boltzmann equation within the relaxation time approximation (RTA)
\begin{eqnarray}
\label{eq:RTABoltzmann}
p^{\mu}\partial_{\mu} f = C[f] = -\frac{p_{\mu}u^{\mu}(x)}{\tau_R} \Big[f-f_{\rm eq}(p_{\mu}\beta^{\mu}(x))\Big]\;, \nonumber \\
\end{eqnarray}
where the coordinate system defined in Minkowski space is $x^{\mu}=(x^{0},\mathbf{x},x^{3})$ with the metric $g_{\mu\nu}=\text{diag}(1,-1,-1,-1)$. In Eq.~\eqref{eq:RTABoltzmann} we denote $\beta^{\mu}(x)=u^{\mu}(x)/T(x)$ with the local rest-frame velocity $u^{\mu}(x)$ determined via the Landau matching condition 
\begin{subequations}
\label{eq:LandauMatching}
\begin{align}
\label{eq:velmatch}
T^{\mu\nu}(x)u_{\nu}(x)&=e(x)u^{\mu}(x)\;,\\
\label{eq:enematch}
 e(x)&=e_{\rm eq}(T(x))\;. 
\end{align}
\end{subequations}
The fluid velocity is defined as a time-like eigenvector ($u^2=+1$) of the energy momentum tensor for on-shell particles
\begin{eqnarray}
\label{eq:Tmn}
T^{\mu\nu}(x)= \langle\,~p^{\mu}p^{\nu}\,\rangle\;, \nonumber
\end{eqnarray}
where we denote the on-shell momentum average of any observable as
\begin{equation} 
\begin{split}
\langle \,\mathcal{O}\,\rangle_X\,=\,\frac{\nu_{\rm eff}}{(2\pi)^4}\,&\int\,\frac{d^4p}{\sqrt{-g(x)}}\,(2\pi)\,\delta(p^2)\,2\theta(p^0)\\
&\times\,\,\mathcal{O}(x^\mu,p^\mu)\,f_X\left(x^\mu,p^\mu\right)\,.
\end{split}
\end{equation}
The effective temperature $T(x)$ entering in Eq.~\eqref{eq:enematch} is determined from the (equilibrium) equation of state $e_{\rm eq}(T)$, by matching the corresponding eigenvalue $e$ to the equilibrium energy density. If not stated otherwise, we will consider an ultra-relativistic system of massless bosons where 
the equilibrium distribution function is given by the Bose-Einstein distribution $f_{eq.}(x)=1/\left(e^{x}-1\right)$, such that
\begin{eqnarray}
e_{\rm eq}(T)=\nu_{\rm eff} \frac{\pi^2}{30} T^{4}\;.
\end{eqnarray} \\
We are interested in longitudinal boost-invariant expanding system and thus, we use the hyperbolic Bjorken coordinates defined in terms of the cartesian coordinates as
\begin{equation}
\label{eq:Milnecoord}
\tau =\sqrt{\left(x^0\right)^2-\left(x^3\right)^2}\,\qquad\,\varsigma=\text{arctanh} \left(x^3/x^0\right)\,,
\end{equation}
so the metric $g_{\mu\nu}=diag(1,-1,-1,-\tau^2)$ and $\sqrt{\,-\,g(x)}=\tau$. Similarly, the four-momentum of a relativistic massless particle is 
\begin{eqnarray}
\label{eq:Milnemom}
p^{\mu}=(p_T \cosh(y),\mathbf{p},p_T \sinh(y))
\end{eqnarray}
with $y=\text{arctanh}\,(p^3/p^0)$ and $p_T=|\mathbf{p}|$. 

In the Bjorken coordinates the Boltzmann equation~\eqref{eq:RTABoltzmann} takes the following form
\begin{eqnarray}
\label{eq:Boltzmann_RTA_Bj}
&&\left[p^{\tau} \partial_{\tau} +p^{i}\partial_{i} + p^{\varsigma} \partial_{\varsigma} \right] f(x,p) =\\  
&& \qquad \qquad -\frac{p_{\mu} u^{\mu}(x)}{\tau_R} \Big[f(x,p)-f_{\rm eq}(p_{\mu}\beta^{\mu}(x))\Big]\;.  \nonumber 
\end{eqnarray}
where 
\begin{eqnarray}
p^{\tau}=p_T \cosh(y-\varsigma)\;, \qquad p^{\varsigma}=\frac{1}{\tau} p_T \sinh(y-\varsigma)\;
\end{eqnarray}
Hereafter we shall use the roman letter $i=x,y$ to denote the summation only over the transverse coordinates. By virtue of the coordinate transformation~\eqref{eq:Milnecoord}, the derivatives w.r.t. $\tau,\varsigma$ in Eq.~(\ref{eq:Boltzmann_RTA_Bj}) are taken at constant $p_T$ and momentum space rapidity $y$.  However, when analyzing the dynamics of a boost invariant medium it is more convenient to work with the (dimensionless) longitudinal momentum variable
\begin{eqnarray}
p_{\varsigma}=-\tau^2 p^{\varsigma}=-\tau p_T \sinh(y-\varsigma)\;.
\end{eqnarray}
By transforming the space-time derivatives according to 
\begin{eqnarray}
\partial_{\tau} &=&\left.\partial_{\tau}\right|_{p_{\varsigma}} - ~~p_T  \sinh(y-\varsigma) \partial_{p_{\varsigma}}\;,\\
\partial_{\varsigma} &=&\left.\partial_{\varsigma}\right|_{p_{\varsigma}} + \tau  p_T \cosh(y-\varsigma) \partial_{p_{\varsigma}}\;, 
\end{eqnarray}
one finds that the two additional terms cancel each other, such that
\begin{eqnarray}
\label{eq:Boltzmann_RTA_Bj2}
&&\left[p^{\tau} \partial_{\tau} +p^{i}\partial_{i} - \frac{p_{\varsigma}}{\tau^2} \partial_{\varsigma} \right] f(x,p) =\\  
&& \qquad \qquad -\frac{p_{\mu} u^{\mu}(x)}{\tau_R} \Big[f(x,p)-f_{\rm eq}(p_{\mu}\beta^{\mu}(x))\Big]\;,  \nonumber 
\end{eqnarray}
which is the form of the equation that we will consider in this work. We note in passing, that it is also common in the literature to express the dynamics in terms of the longitudinal momentum in the local rest frame
\begin{eqnarray}
p^{\|}= \tau p^{\varsigma}=p_T \sinh(y-\varsigma)\;.
\end{eqnarray}
In this case only the derivative w.r.t. the longitudinal rapidity is affected by the transformation
 \begin{eqnarray}
\partial_{\varsigma} &=&\left.\partial_{\varsigma}\right|_{p^{\|}} -  p_T  \cosh(y-\varsigma)\partial_{p^{\|}}\;, 
\end{eqnarray}
and the Boltzmann equation takes the form (see e.g. \cite{Mueller:1999pi})
\begin{eqnarray}
&&\left[p^{\tau} \partial_{\tau} +p^{i}\partial_{i} + \frac{p^{\|}}{\tau} \partial_{\varsigma} - p^{\tau} \frac{p^{\|}}{\tau} \partial_{p^{\|}} \right] f(x,p) =\\  
&& \qquad \qquad -\frac{p_{\mu} u^{\mu}(x)}{\tau_R} \Big[f(x,p)-f_{\rm eq}(p_{\mu}\beta^{\mu}(x))\Big]\;.  \nonumber 
\end{eqnarray}

\section{Evolution of boost invariant homogenous background}
\label{sec:back}

During the early pre-equilibrium stage of a heavy-ion collision, which lasts about $1 \rm{fm}/c$, the non-equilibrium plasma is subject to a rapid longitudinal expansion. Conversely, in the transverse plane the plasma is initially created at rest, and the transverse expansion only builds up in response to local energy-density gradients on a time scale $\tau \sim R$, where $R$ denotes the system size. Due to this separation of scales, it is a reasonable assumption to neglect the transverse expansion during the pre-equilibrium stage, and first consider an idealized situation of Bjorken flow, where the system is longitudinally boost-invariant, parity invariant under spatial reflexions along the longitudinal beam line and  azimuthally symmetric and translationally invariant in the transverse plane. Space-time dependent variations can subsequently be addressed by studying small deviations from this average background behavior, and will be discussed in Sec.~\ref{sec:EMpert}.

In this section we discuss the space-time evolution of an expanding background undergoing Bjorken expansion, such that the aforementioned symmetries constrain the functional form of the distribution function for the background to be of the form
\begin{eqnarray}
f(x,p)=f_{BG}(\tau,p_T,|p_{\varsigma}|)\;.
\end{eqnarray}
The energy momentum tensor has only non-vanishing diagonal components, i.e.,
\begin{eqnarray}
\label{eq:tmn_back}
T^{\mu\nu}_{BG}=diag(e,p_T,p_T,p_L/\tau^2)\;,
\end{eqnarray}
with the energy density ($e$), transverse and longitudinal pressures ($p_{T/L}$) determined by
\begin{subequations}
\label{eq:tmncomp}
\begin{align}
e&=\,T^{\tau\tau}_{BG}=\bigl\langle\,\left(\,p^\tau\,\right)^2\,\bigr\rangle_{BG}\;, \\
p_T&=T^{xx}_{BG}=T^{yy}_{BG}=\frac{1}{2} \bigl\langle\,\left(\,p_T\,\right)^2\,\bigr\rangle_{BG}\;, \\
p_L&=T^{\varsigma\varsigma}_{BG}=\biggl\langle\,\left(\frac{p_\varsigma}{\tau}\,\right)^2\biggr\rangle_{BG}\;.
\end{align}
\end{subequations}
Due to scale invariance, the previous expressions automatically satisfy the tracelessness condition $e=2\,p_T+p_L$. Based on the explicit form of $T^{\mu\nu}_{BG}$ the Landau matching condition in Eq.~(\ref{eq:LandauMatching}) becomes trivial with
\begin{subequations}
\begin{align}
u^{\mu}&=(u^{\tau},u^{i},u^{\varsigma})=(1,0,0,0)\;, \\
e&=e_{\rm eq}\;,
\end{align}
\end{subequations}
and the kinetic equation for the evolution of the background distribution takes the familiar form
\begin{eqnarray}
\label{eq:BG_RTA}
&&\tau\partial_{\tau} f_{BG}(\tau,p_T,|p_{\varsigma}|) = \\
 &&-\frac{\tau}{\tau_{R}} \left[f_{BG}\Big(\tau,p_T,|p_{\varsigma}|\Big) - f_{\rm eq}\Big( \frac{p^\tau}{T(\tau)}\Big)\right]\;. \nonumber
\end{eqnarray}
where $p^\tau=\sqrt{p_T^2\,+\,(p_\varsigma/\tau)^2}$ according to the on-shell mass condition.

\subsection{Evolution equations for moments}
Several strategies have been explored in the literature to solve Eq.~(\ref{eq:BG_RTA})~\cite{Groot:1980}. Here we follow Grad's original approach~\cite{Grad} where instead of finding solutions for the phase-space distribution function $f(x,p)$ we study the dynamics of its moments. The latter are directly connected to macroscopic observables as we indicate below. 

We consider the following moments 
\begin{eqnarray}
\label{eq:CMom}
C_{l}^{m}(\tau) &=& \nu_{\rm eff} \int \frac{dp_{\varsigma}}{(2\pi)}  \int \frac{d^2{\bf p}}{(2\pi)^2} \tau^{1/3} \sqrt{p_T^2+(p_{\varsigma}/\tau)^2}  \nonumber \\
&& \qquad Y_{l}^{m}(\phi_p,\theta_{p})~f_{BG}(\tau,p_T,|p_{\varsigma}|) \;, 
\end{eqnarray}
where the angles are defined as $\cos\theta_p=p_\varsigma/(\tau p^\tau)$, $\tan\phi_p=p^1/p^2$ in the co-moving coordinates and $Y_{l}^{m}$ denote the spherical harmonics
\begin{eqnarray}
Y_{l}^{m}(\phi,\theta)&=&y_{l}^{m}~P_{l}^{m}(\cos(\theta))e^{im\phi}\;, 
\end{eqnarray}
with normalization
\begin{eqnarray}
y_{l}^{m}&=&\sqrt{\frac{(2l+1)(l-m)!}{4\pi (l+m)!}}\;.
\end{eqnarray}
Non-vanishing components of the background energy-momentum tensor in~\eqref{eq:tmn_back} are related with the moments $C_l^m$ in \eqref{eq:CMom} as follows
\begin{eqnarray}
C^{0}_{0}(\tau)&=&\sqrt{\frac{1}{4\pi}} \tau^{4/3} ~e(\tau)\;, \\
C^{0}_{2}(\tau)&=&\sqrt{\frac{5}{16\pi}} \tau^{4/3} \Big[3p_L(\tau) -e(\tau)\Big]\;.
\end{eqnarray}
Specifcially, if the background distribution function is in thermal equilibrium, one has
\begin{eqnarray}
\label{eq:CeqVals}
\left.C_{l}^{m}\right|_{\rm eq}(\tau)=\frac{\tau^{4/3}e(\tau)}{\sqrt{4\pi}} \delta_{l0}\delta^{m0}\;,
\end{eqnarray}
where the Landau matching conditions~\eqref{eq:LandauMatching} was enforced.  
Now the evolution equations of those moments is simply obtained by taking the explicit time derivative in its definition. After some careful algebra and using a series of identities of the spherical harmonics (see App.~\ref{sec:SphericalHarmonicIdentities}), the evolution equation for the moments $C_l^m$ takes the following form
\begin{eqnarray}
\label{eq:BGClmEvo}
&&\tau \partial_{\tau} C_{l}^{m} = \nonumber \\
&& b_{l,-2}^{m}  C_{l-2}^{m} + b_{l,0}^{m}  C_{l}^{m} + b_{l,+2}^{m}  C_{l+2}^{m} \nonumber \\
&& - \frac{\tau}{\tau_R} \left( C_{l}^{m} - \left.C_{l}^{m}\right|_{\rm eq}\right)\;,
\end{eqnarray}
where the coefficients $b_{l,-2}^m$, $b_{l,0}^{m}$ and $b_{l,+2}^{m}$ are given by
\begin{eqnarray}
\label{eq:BCoefficientsDef}
b_{l,-2}^{m} &=& \frac{l+2}{2l+1} \sqrt{\frac{\Gamma(l-m+1) \Gamma(l+m+1) }{(2l+1)(2l-3)\Gamma(l-m-1) \Gamma(l+m-1)}} \nonumber \\
b_{l,0}^{m} &=& - \frac{5}{3} \frac{l (l+1) - 3 m^2}{4l(l+1)-3} \\
b_{l,+2}^{m} &=& - \frac{l-1}{2l+3} \sqrt{\frac{\Gamma(l-m+3) \Gamma(l+m+3) }{(2l+1)(2l+5)\Gamma(l-m+1) \Gamma(l+m+1)}} \nonumber
\end{eqnarray}
We note that in Eq.~(\ref{eq:BGClmEvo}) only the moments $l$ and $l\pm 2$ are coupled reflecting the parity symmetry of the background. Now due to the azimuthal symmetry of the background, the evolution of moments with different $m$ are not coupled and only moments with $m=0$ are non-vanishing. Furthermore, Eqs.~\eqref{eq:BGClmEvo} present interesting mathematical resurgent properties which were discussed extensively in Refs.~\cite{Behtash:2018moe,Behtash:2019txb,Behtash:2019qtk}. 

\subsection{Initial conditions} 
Since at early times $\tau \ll \tau_R$ the system is unable to sustain sizeable longitudinal momenta, the physically relevant initial conditions for the phase-space distribution are naturally of a form where $p_\| \ll p_T$. Indeed, previous works~\cite{Strickland:2018ayk,Behtash:2018moe,Behtash:2019qtk,Dash:2020zqx,Blaizot:2017lht,Blaizot:2017ucy,Blaizot:2019scw} have shown that the extreme limit where the (longitudinal) support of the phase-space distribution shrinks to a Dirac delta function corresponds to a non-equilibrium attractor of the kinetic equation. We will therefore consider precisely this case and choose the initial phase-space distribution as
\begin{eqnarray}
\label{eq:inidistfunc}
f_{BG}(\tau_0,p_T,p_\varsigma) = \frac{(2\pi)^3}{\nu_{\rm eff}}~\delta(p_{\varsigma}) \frac{dN_{0}}{d\varsigma d^2\pt d^2\xt}\;,
\end{eqnarray}
where the normalization chosen such that the initial energy density per unit rapidity remains constant, i.e.
\begin{eqnarray}
\frac{dE_{0}}{d\varsigma d^2\xt}=\lim_{\tau_0\to0} \tau_0 e(\tau_0) = \left( \tau e \right)_0 = const\;.
\end{eqnarray}
By construction, the initial condition in Eq.~(\ref{eq:inidistfunc}) fixes the initial values of the moments $C_{l}^{m}$ at $\tau_0$, i.e.,
\begin{eqnarray}
\label{eq:CMomInitialCond}
C_{l}^{m}(\tau_0)= \tau_0^{1/3}~\left( \tau e\right)_0~y_{l}^{m}~P_{l}^{m}(0) \delta^{m0}
\end{eqnarray}
where
\begin{eqnarray}
P_{l}^{m}(0)=\frac{2^{m} \sqrt{\pi}}{\Gamma\left(\frac{1-l-m}{2}\right)\Gamma\left(1-\frac{m-l}{2}\right)}\;,
\end{eqnarray}
such that for $m=0$ one has
\begin{eqnarray}
P_{l}^{0}(0)=\frac{\sin(\pi/2(l+1))}{\sqrt{\pi}} \frac{\Gamma(1/2+l/2)}{\Gamma(1+l/2)}\;.
\end{eqnarray}
Interestingly, one finds that at the level of relaxation time approximation, the shape of the azimuthally symmetric momentum distribution $\frac{dN}{d\varsigma d^2\pt d^2\xt}$ is completely irrelevant for the dynamics. Noteably, this is in sharp contrast to more realistic description in Yang-Mills kinetic theory \cite{Kurkela:2015qoa,Kurkela:2018wud,Kurkela:2018vqr,Kurkela:2018xxd}, where different processes (e.g. inelastic and elastic interactions) exhibit different parametric dependencies on the momenta (see e.g. \cite{Schlichting:2019abc,Dusling:2009df}). \\

\subsection{Evolution of energy momentum tensor for constant relaxation time ($\tau_R=const$)}

\begin{figure}[t!]
\label{fig:1}
\centering
\includegraphics[width=0.5\textwidth]{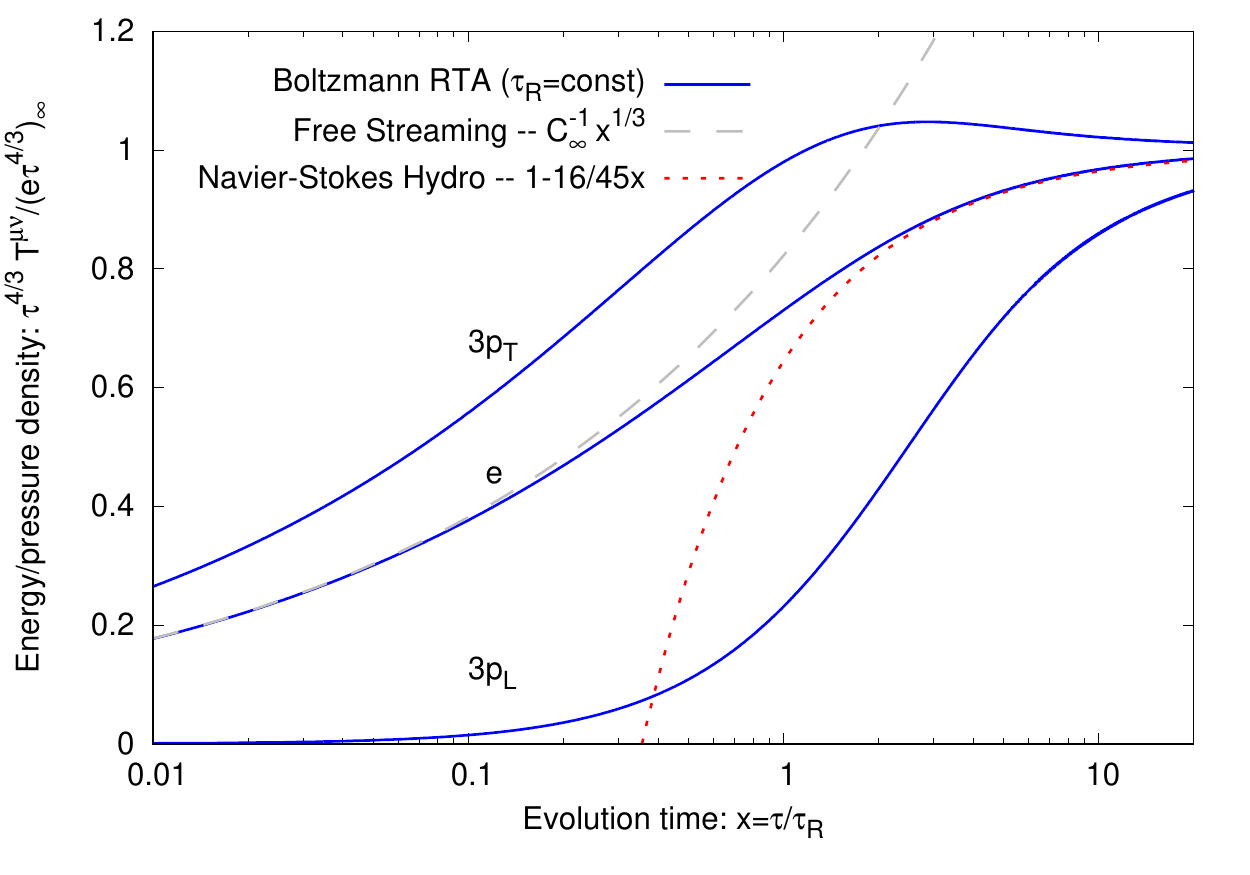}
\includegraphics[width=0.5\textwidth]{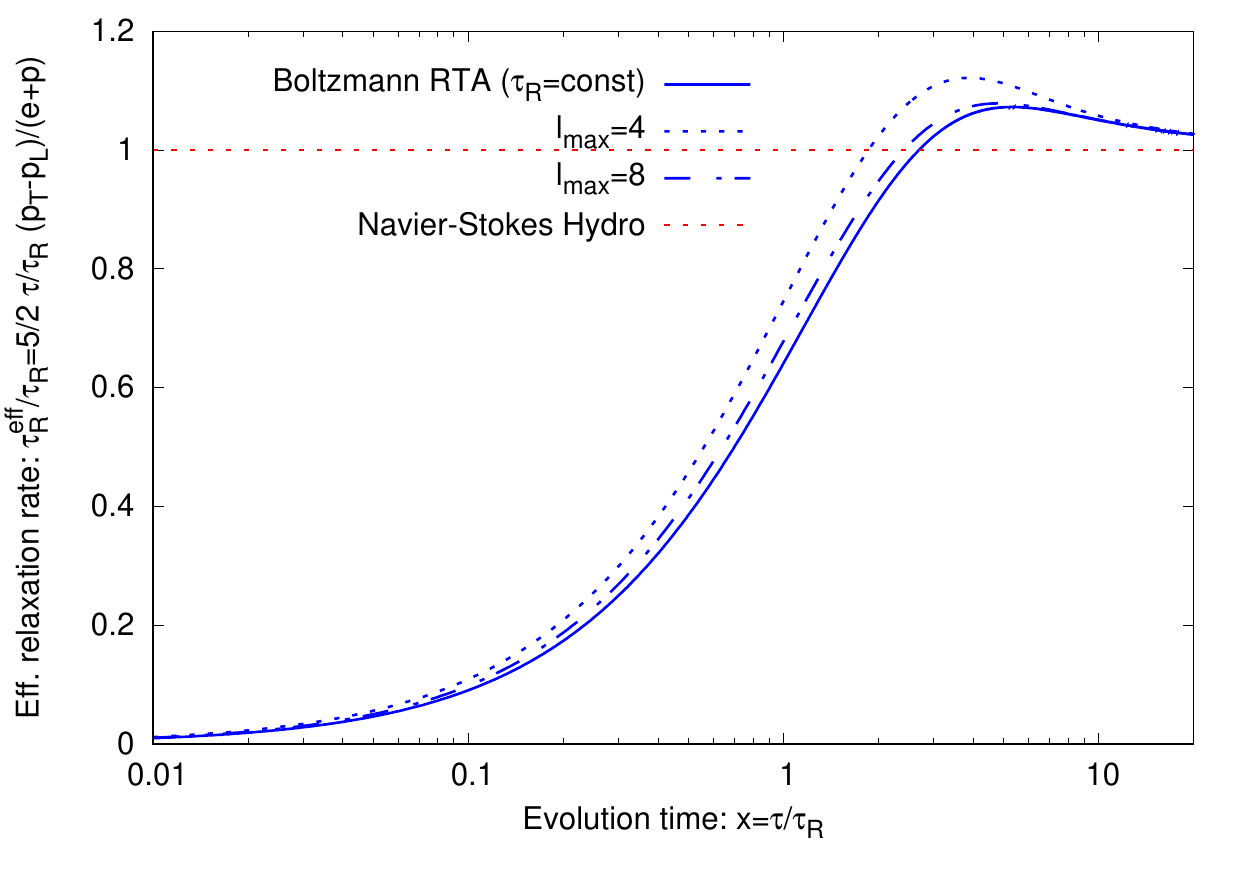}
\caption{\label{fig:BGConst} (top) Evolution of energy-momentum tensor for const. relaxation time. (bottom) Effective relaxation time determined from the appropriate ratio of inverse Reynolds and the Knudsen number (see text).}
\end{figure}

We first analyze the evolution of the energy momentum tensor for a constant relaxation time $\tau_{R}=const$. By truncating the evolution equations up to a certain value of $l<l_{\rm max}$, setting $C_{l}^{m}=\left. C_{l}^{m} \right|_{\rm eq}=0$, we can obtain a numerical solution of the coupled set of evolution equation as a function of the dimensionless evolution time variable $\tau/\tau_{R}$. Noteably the truncation scheme converges rapidly except for very small values of $\tau/\tau_R \ll 1$ where higher moments play an important role. However, for the typical regime of interest no visible deviations of the results for the lowest $l \leq 2$ moments occur for $l_{max} \gtrsim 32$, and if not stated otherwise we use $l_{max} =512$ to produce the figures.  

\subsubsection{Comparison with hydrodynamics}
Our results are compactly summarized in Fig.~\ref{fig:BGConst}, which shows the non-equilibrium attractor solution for the evolution of the energy and pressure densities. For the Bjorken flow we can also compare the numerical solution of the Boltzmann equation in relaxation time approximation to the truncation in relativistic viscous hydrodynamics. Starting from the conservation equation 
\begin{eqnarray}
\tau \partial_{\tau} e= - e-p_L\;,
\end{eqnarray}
the longitudinal and transverse pressure within the hydrodynamic approach are given by 
\begin{eqnarray}
p_L=p+\pi^\varsigma_{\hspace{.1cm}\varsigma}\;, \qquad p_T=p-\frac{\pi^\varsigma_{\hspace{.1cm}\varsigma}}{2}
\end{eqnarray}
where $\pi^\varsigma_{\hspace{.1cm}\varsigma}$ is the only independent component of the shear viscous tensor. Now $\pi^\varsigma_{\hspace{.1cm}\varsigma}$ is expanded up to second order in the gradient hydrodynamical expansion as~\cite{Baier:2007ix}
\begin{equation}
\label{eq:gradexp}
\pi^\varsigma_{\hspace{.1cm}\varsigma}=-\frac{4}{3}\frac{\eta}{\tau}+\frac{8}{9}\frac{1}{\tau^2}\left(\lambda_1-\eta\tau_\pi\right)\,,
\end{equation}
where for the system under consideration the equation of state and transport coefficients in the RTA approximation are determined by ~\cite{Teaney:2013gca,Jaiswal:2013npa,Blaizot:2017lht,Behtash:2019txb}
\begin{equation}
\label{eq:RTArel}
\begin{split}
p=e/3\;, &\qquad \eta/s=\frac{\tau_R T}{5}\;, \qquad Ts=e+p\;,\\
&\tau_\pi=\tau_R\;,\qquad \lambda_1=\frac{3}{35}\tau_R^2\,Ts
\end{split}
\end{equation}
such that $\eta/\tau=\frac{4}{15} \frac{\tau_R}{\tau}\,e$. Hence the asymptotic solution for the energy density up to second order within the gradient hydrodynamic expansion (or alternatively when $\tau/\tau_R\gg 1$) takes the form 
\begin{equation}
\label{eq:ConstantAsymptotic}
\tau^{4/3} e(\tau) \approx\left(\tau^{4/3} e \right)_{\infty} \left( 1- \frac{16}{45} \frac{\tau_{R}}{\tau}-\frac{64}{14175}\left(\frac{\tau_R}{\tau}\right)^2\right)\;,
\end{equation}
with the asymptotic integration constant $ \left(\tau^{4/3} e \right)_{\infty}$ determined as
\begin{eqnarray}
\lim_{\tau \to  \infty} \tau^{4/3} e(\tau)  = \left(\tau^{4/3} e \right)_{\infty} \simeq 1.2~\tau_{R}^{1/3}~\left(\tau e \right)_{0}
\end{eqnarray}
from the numerical solution of the Boltzmann equation in relaxation time approximation. By comparing the different curves in Fig.~\ref{fig:BGConst} one observes that viscous hydrodynamics starts to describe the evolution of the energy momentum tensor around evolution times $\tau/\tau_{R} \sim 2-3$. Since including the second order correction does not significantly improve the agreement in Fig.~\ref{fig:BGConst}, we only present the Navier-Stokes limit, i.e. the leading term in the asymptotic expansion in $\tau_{R}/\tau$ in Eq.~(\ref{eq:ConstantAsymptotic}).

Noteably, one can also define an effective ratio between the inverse Reynolds and the Knudsen number as follows
\begin{eqnarray}
\label{eq:ReKn}
\frac{Re^{-1}}{Kn}= \,4\,\frac{\tau}{\tau_R}\,\left(\frac{|\pi^{\varsigma}_{\varsigma}|}{e+p}\right)\;,
\end{eqnarray}
with $Kn=\tau_R/\tau$  being the Knudsen number and the inverse Reynolds number $Re^{-1}$ is 
\begin{equation}
\label{eq:Kn}
Re^{-1}\sim\frac{|\pi^{\varsigma}_{\varsigma}|}{p}=\frac{2}{3}\frac{|p_L-p_T|}{p}\,.
\end{equation}
Near equilibrium where the gradient expansion~\eqref{eq:gradexp} holds, one can express the inverse Reynolds number as
\begin{equation}
Re^{-1}=\frac{16}{3}\frac{\eta}{s}\,Kn\,+\,\mathcal{O}(Kn^2)\,.
\end{equation}
While the gradient expansion~\eqref{eq:gradexp} breaks down at early times, where the Knudsen number $Kn \gg 1$ , the ratio in Eq.~(\ref{eq:ReKn}) can still be used to quantify the evolution. Specficially, for the case of a constant relaxation time, one can use Eq.~(\ref{eq:RTArel}) to define  an effective relaxation rate $\tau_R^{eff}$ away from equilibrium
\begin{equation}
\label{eq:efftaur}
\frac{\tau_R^{eff}}{\tau_R}=\frac{15}{16}\,\left|\frac{Re^{-1}}{Kn}\right|\,.
\end{equation}
which is constructed such that at late times the ratio $\frac{\tau_R^{eff}}{\tau_R}$ in Eq.~\eqref{eq:efftaur} converges to unity as expected. 

Numerical results for $\frac{\tau_R^{eff}}{\tau_R}$ shown in the bottom panel of Fig.~\ref{fig:BGConst} indicate that at early times where the Knudsen number $Kn \gg 1$, the effective relaxation rate is significantly reduced by the inclusion of higher order dynamical moments $C_{l}^{m}$. By explicitly comparing different truncations $(l_{\rm max}=4,8,\cdots)$ of the infinite hierarchy of moment equations, one also observes a rapid convergence in the sense that the evolution of the low order moments becomes increasingly insensitive to the higher order moments.


\subsection{Evolution of energy momentum tensor for conformal system ($T(\tau) \tau_R=const$)}
\begin{figure}[t!]
\centering
\includegraphics[width=0.5\textwidth]{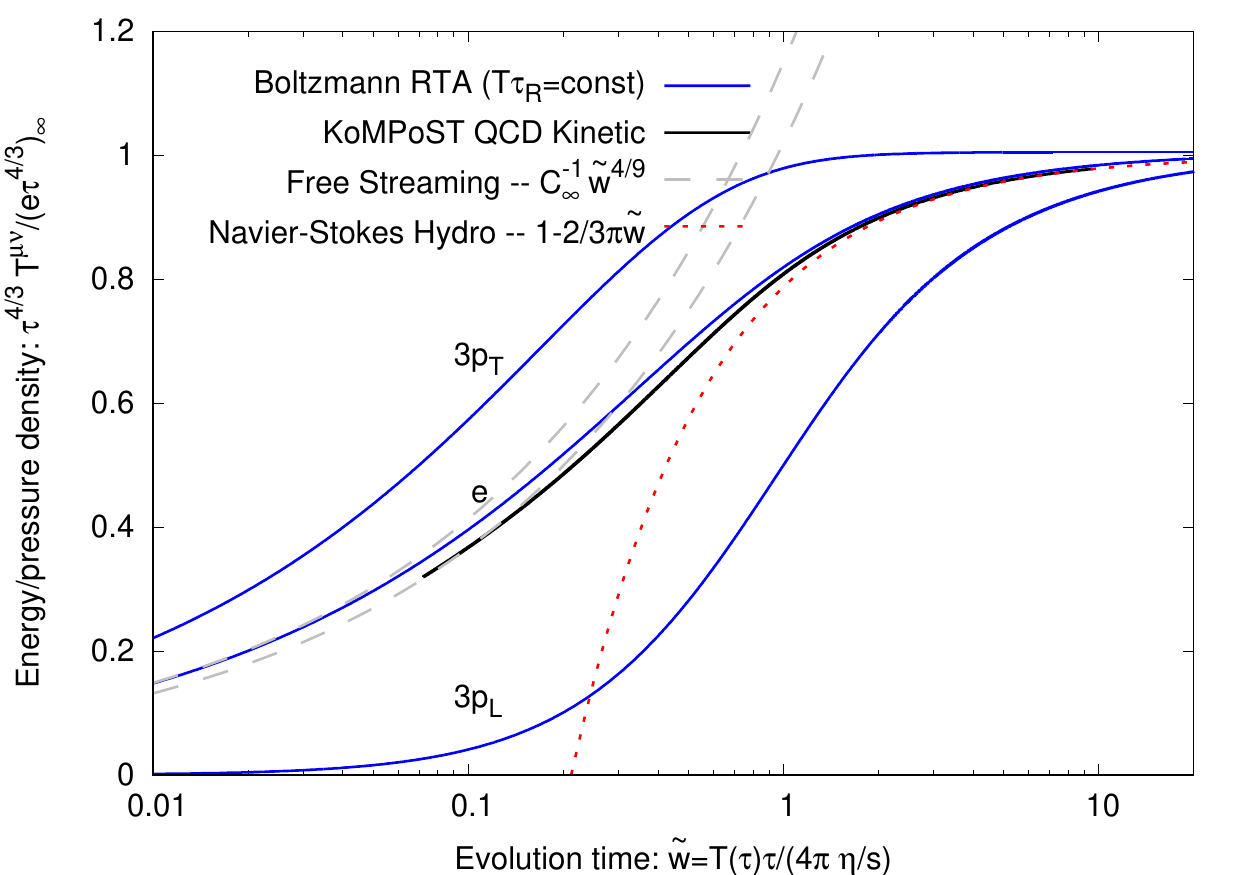}
\includegraphics[width=0.5\textwidth]{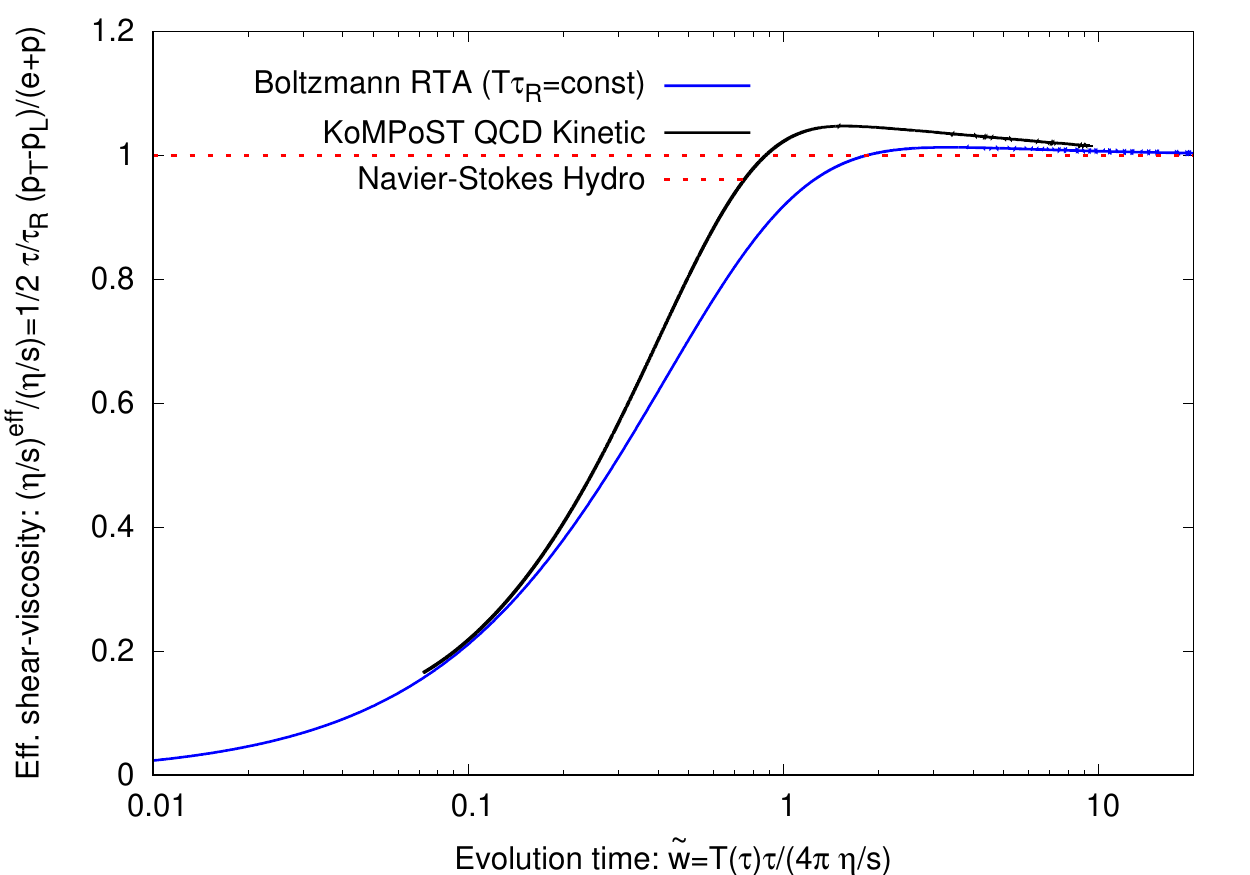}
\caption{\label{fig:BGConformal} (top) Evolution of energy-momentum tensor in the relaxation time approximation with constant $\eta/s$, compared to Yang-Mills kinetic theory (\kompost) \cite{Kurkela:2018vqr,Kurkela:2018wud}. (bottom) Evolution of the effective viscosity.}
\label{fig:2}
\end{figure}

We now investigate the more commonly studied case where the relaxation time is chosen inversely proportional to the eff. temperature, i.e. $\tau_{R}T(\tau)=5(\eta/s)=const$, as is appropriate for a conformal system~ \cite{Blaizot:2019scw,Blaizot:2017ucy,Strickland:2018ayk,Heller:2016rtz,Florkowski:2013lza}. We conveniently introduce the dimensionless time-like variable
\begin{eqnarray}
\label{eq:xEvoDef}
x=\frac{\tau}{\tau_R} = \frac{T(\tau)\tau}{5\eta/s} =\frac{4\pi}{5} \tilde{w}\;.
\end{eqnarray}
which one can also identify as the inverse Knudsen number $x\equiv Kn^{-1}$.  Since for the conformal system, the relaxation time depends on the temperature of the system as $\tau_R=5\eta/sT(\tau)$, it is convenient to perform a change of variables to express
\begin{eqnarray}
\tau\partial_{\tau}= a(x) x \partial_{x}\;,
\end{eqnarray}
where we defined the scale-factor
\begin{eqnarray}
a(x)=\left[ \frac{2}{3} + \frac{1}{4} \left( b_{0,0}^{0} + b_{0,+2}^{0} \frac{C_{2}^{0}(x)}{C_{0}^{0}(x)}\right) \right]\;,
\end{eqnarray}
and used the equation of motion for the energy density along with the fact that for an ultra-relativistic system
$\frac{dT}{T} = \frac{de}{4e}$. The evolution equation for the moments can be then re-cast into the form
\begin{eqnarray}
a(x) x \partial_{x} C_{l}^{m} &=& b_{l,-2}^{m}  C_{l-2}^{m} + b_{l,0}^{m}  C_{l}^{m} + b_{l,+2}^{m}  C_{l+2}^{m} \nonumber \\
&&  - x \left( C_{l}^{m} - \left.C_{l}^{m}\right|_{\rm eq}\right)\;. 
\end{eqnarray}
indicating that for a given initial condition the non-equilibrium evolution of the system is uniquely determined by the conformal scaling variable $x$.

Numerical results for the evolution of the background energy-momentum tensor are presented in Fig.~\ref{fig:BGConformal}, where we also compare to the results for Yang-Mills kinetic theory obtained in \cite{Kurkela:2018vqr,Kurkela:2018wud}. Clearly the overall behavior of the curves is quite similar, showing a smooth transition from an approximate free-streaming behavior at early times towards the universal hydrodynamic behavior
\begin{eqnarray}
\tau^{4/3} e(\tau) = \left(\tau^{4/3} e \right)_{\infty} \left( 1- \frac{8}{3} \frac{\eta/s}{T(\tau) \tau} + \cdots \right)\;.
\end{eqnarray}
at late times. Similarly, to our previous discussion, this behavior can also be analyzed in terms of an effective shear viscosity over entropy ratio in far from equilibrium regime, which for the conformal system is given by  
\begin{equation}
\label{eq:effshear}
\frac{\left(\eta/s\right)_{eff}}{\eta/s}=\frac{3}{16}\,\left|\frac{Re^{-1}}{Kn}\right|\,,
\end{equation} 
which is again constructed such that $\frac{\left(\eta/s\right)_{eff}}{\eta/s}$ approaches unity in the limit $\tau/\tau_{R} \gg 1$. Even though this ratio depicted in the lower panel of Fig.~\ref{fig:BGConformal} magnifies the differences between the results for Yang-Mills kinetic theory \cite{Kurkela:2018vqr,Kurkela:2018wud} and the relaxation time approximation, the overall differences are at most at the $10\%$ level for intermediate times $\tilde{w} \sim 1$.

Nevertheless, the fact that the approach towards visc. hydrodynamics is different for Yang-Mills kinetic theory and RTA results in a mismatch in the ratios of the initial energy density to final energy density. Since the early time behavior is governed by free-streaming, one finds that  $e(\tau) \propto 1/\tau$ for $ x \ll 1$ such that the attractor curve can be parametrized as
\begin{eqnarray}
\frac{\tau_0^{4/3} e_{0}}{\left(\tau^{4/3} e \right)_{\infty} } \stackrel{x\ll1}{=} \frac{1}{C_{\infty}}  \left( \frac{T(\tau_{0}) \tau_{0}}{4\pi \eta/s}\right)^{4/9}\;.
\end{eqnarray}
By inverting this relation, one then obtains the energy density at late times as \cite{Giacalone:2019ldn}
\begin{eqnarray}
\left(\tau^{4/3} e \right)_{\infty}=C_{\infty}~\left(\frac{4\pi \eta/s}{T(\tau_{0}) \tau_{0}} \right)^{4/9}~(e \tau)_{0}\;.
\end{eqnarray}
Specifically, for the conformal relaxation time approximation, we find $C_{\infty} \approx 0.9$ whereas for the Yang-Mills kinetic theory results of \cite{Kurkela:2018vqr,Kurkela:2018wud} the pre-factor $C_{\infty}\approx 1$ is about ten percent larger \cite{Giacalone:2019ldn}.

\section{Energy momentum perturbations around Bjorken flow}
\label{sec:EMpert}

So far we have addressed the non-equilibrium evolution of the average boost invariant and homogenous background. We will now consider the propagation of linearized perturbations, sourced by (small) space-time dependent deviations of the initial energy momentum tensor from its (local) average. By linearizing the kinetic equation around the boost invariant and homogenous background  one finds an evolution equation for the perturbation of the distribution function $\delta f$, i.e., 
\begin{widetext}
\begin{eqnarray}
\label{eq:linBolGen}
 \left[ p^{\tau} \partial_{\tau} + p^{i} \partial_{i} -\frac{p_{\varsigma}}{\tau^2} \partial_{\varsigma} \right] \delta f(x,p) &=& - \frac{p^{\tau}}{\tau_R} \delta f(x,p) - \frac{p_{\mu} \delta u^{\mu}(x)}{\tau_R} \left[ \left(f_{BG}(\tau,p_T,|p_\varsigma|)-f_{\rm eq}  \Big( \frac{p^{\tau}}{T(\tau)} \Big)\right)  - \frac{p^{\tau}}{T(\tau)} f'_{\rm eq} \Big( \frac{p^{\tau}}{T(\tau)} \Big) \right] \nonumber \\
&&- \frac{p^\tau}{\tau_R} \frac{\delta T(x)}{T(\tau)} \left[\frac{p^{\tau}}{T(\tau)}   f'_{\rm eq} \Big( \frac{p^{\tau}}{T(\tau)} \Big) -\frac{T(\tau)}{\tau_R} \frac{\partial \tau_R}{\partial T} \left(f_{BG}(\tau,p_T,|p_\varsigma|)- f_{\rm eq}  \Big( \frac{p^{\tau}}{T(\tau)} \Big) \right) \right]\;.
\end{eqnarray}
\end{widetext}
where $f'_{\rm eq}(x)=d f_{\rm eq}(x)/dx$ denotes the derivative of the equilibrium distribution. In the above expression, the change in the rest-frame velocity $\delta u^{\mu}(x)$ and local equilibrium temperature $\delta T(x)$ are to be determined self-consistently from the (linearized) Landau matching condition. Starting from the linearized perturbations of the energy momentum tensor 
\begin{eqnarray}
&&\delta T^{\mu\nu}(x)=\,\langle\,p^\mu\,p^\mu\,\rangle_{\delta f}
\end{eqnarray}
the change in the rest-frame velocity $\delta u^\mu$ and energy density in the local rest-frame $\delta e$ are determined from the linearized eigenvalue equation
\begin{eqnarray}
u_{\mu} \delta T^{\mu\nu} + \delta u_{\mu} T^{\mu\nu} = \delta e u^{\nu} + e \delta u^{\nu}\;, 
\end{eqnarray}
with
\begin{eqnarray}
u_{\mu} \delta u^{\mu}=0\;.
\end{eqnarray}
By using the leading order solution $u_{\mu} T^{\mu\nu} = e u^{\nu}$ of the eigenvalue problem, one reads off
\begin{subequations}
\label{eq:linLandmatch}
\begin{align}
&\delta e = u_{\mu} \delta T^{\mu\nu} u_{\nu}=\delta T^{\tau\tau}\;, \\
&\delta u^{\tau}=0\;, \quad
\delta u^{i} =\frac{\delta T^{\tau i}}{e+p_T}\;,  \quad
\delta u^{\varsigma} =\frac{\delta T^{\tau \varsigma}}{e+p_L}\;. 
\end{align}
\end{subequations}
Our strategy to determine the evolution of energy and momentum perturbations then consist in determining all the coefficients on the r.h.s. of Eq.~(\ref{eq:linBolGen}) based on certain moments of the perturbation of the phase-space distribution function. In the following section we shall explain this procedure in detail.

\subsection{Evolution equations for energy-momentum perturbations in the transverse plane}

We will from now on restrict our attention to energy-momentum pertubations in the transverse plane, i.e. we will only consider variations in the transverse coordinates $\xt$. Since we are considering the evolution of linearized perturbations on top of a (transversely) homogenous background, it is natural to express them in a Fourier basis such that for each $\kt$-mode 
\begin{eqnarray}
\delta f(\tau,{\bf x},{\bf p},|p_\varsigma|)= \int\,\frac{d^2{\bf k}}{(2\pi)^2}\delta f_{\bf k}(\tau,\pt,|p_{\varsigma}|)~e^{i {\bf k}\cdot {\bf x}}\;,
\end{eqnarray}
where we denote $\delta f_{\bf k}(\tau,\pt,|p_{\varsigma}|)\equiv \delta f(\tau,{\bf k},\pt,|p_{\varsigma}|)$. In our approach the $\mathbb{Z}_2$ subgroup symmetry of reflections along the beam axis is not broken at the level of the perturbations so the longitudinal rest frame velocity $\delta u^{\varsigma}$ vanishes identically. The transverse flow velocity can be decomposed in the components parallel ($\delta u_{\bf k}^{\|}(\tau)$) and perpendicular ($\delta u_{\bf k}^{\bot}(\tau)$) to the wave vector ${\bf k}$. Hence the perturbations to the thermodynamic fields take the following form
\begin{subequations}
\begin{align}
 \delta T(\tau,{\bf x})&= \int\,\frac{d^2{\bf k}}{(2\pi)^2}\delta T_{\bf k}(\tau) e^{i{\bf k}\cdot{\bf x}}\;,\\
\delta u^{i}(\tau,{\bf x})&= \int\frac{d^2{\bf k}}{(2\pi)^2}\,\frac{{\bf k}^{j}}{|\bf k|}~\left[  \delta u_{\bf k}^{\|}(\tau) \delta^{ji} +  \delta u_{\bf k}^{\bot}(\tau) \epsilon^{ji}   \right]  \,e^{i{\bf k}\cdot{\bf x}}  \;, \\ 
&\delta u^{\tau}(x)=0\;, \qquad \delta u^{\varsigma}(x)=0\;, \label{eq:uconst}
\end{align}
\end{subequations}
where for physical perturbations, the reality conditions of the perturbations imply $\delta T_{\bf k}(\tau)=\delta T^*_{-\bf k}(\tau)$ and $\delta u^{i}_{\bf k}(\tau)=\delta u^{i, *}_{-\bf k}(\tau)$. Denoting 
\begin{eqnarray}
\frac{{\bf k} \cdot {\bf p}}{|{\bf k}|~p^{\tau}}  &=& \delta^{ij} \frac{\kt^{i} \pt^{j}}{|{\bf k}|~p^{\tau}} = \cos(\phi_{\bf pk}) \sin(\theta_{\bf p})\;, \\
\frac{{\bf k} \times {\bf p}}{|{\bf k}|~p^{\tau}}  &=& \epsilon^{ij} \frac{\kt^{i}\pt^{j}}{|{\bf k}|~p^{\tau}} = \sin(\phi_{\bf pk}) \sin(\theta_{\bf p}) \;,
\end{eqnarray}
with $\phi_{\bf pk}=\phi_{\bf p}-\phi_{\bf k}$ being the angle between ${\bf p}$ and ${\bf k}$ in the transverse plane and $\sin\theta_{\bf p}=|\mathbf{p}|/p^\tau$  and inserting the above expressions into the linearized kinetic equation (\ref{eq:linBolGen}) one then finds
\begin{widetext}
\begin{align}
\label{eq:pertboltz}
&\left[\tau\,\partial_{\tau} + i\,(|{\bf k}| \tau)\,\frac{{\bf k} \cdot {\bf p}}{|{\bf k}|\,p^\tau} \right]  \delta f_{\kt}(\tau,\pt,|p_{\varsigma}|)=
-\left(\frac{\tau}{\tau_R} \right)\,\delta f_{\kt}(\tau,\pt,|p_{\varsigma}|)\,\nonumber \\
& \qquad -\left(\frac{\tau}{\tau_R}\right)\,\left[\frac{\delta T_{\bf k}(\tau)}{T(\tau)} + \delta u^\parallel_{\bf k}(\tau)\,\frac{{\bf k} \cdot {\bf p}}{|{\bf k}|\,p^\tau} + \delta u^\perp_{\bf k}(\tau)\,\frac{{\bf k}\times{\bf p}}{|{\bf k}|\,p^\tau}\right] ~\frac{p^\tau}{T(\tau)}   f'_{\rm eq} \Big( \frac{p^{\tau}}{T(\tau)} \Big) \, \nonumber \\
& \qquad +\left(\frac{\tau}{\tau_R}\right)\,\left[\delta u^\parallel_{\bf k}(\tau)\,\frac{{\bf k} \cdot {\bf p}}{|{\bf k}|\,p^\tau}+\delta u^\perp_{\bf k}(\tau)\,\frac{{\bf k}\times{\bf p}}{|{\bf k}|\,p^\tau} + \frac{\delta T_{\bf k}(\tau)}{T(\tau)}\frac{T(\tau)}{\tau_R} \frac{\partial \tau_R}{\partial T} \right] \left( f_{BG}(\tau,|{\bf p}|,|p_\varsigma|)-f_{\rm eq}  \left( \frac{p^{\tau}}{T(\tau)}\right) \right)\;,
\end{align}
\end{widetext}
where the terms on the left hand side describe free-streaming, the terms in the first line correspond to the relaxation of the perturbation, the terms in the second line describe the change of the equilibrium distribution due to the perturbations and the terms in the last line describe the change in the relaxation of the non-equilibrium background due to perturbations.

\subsection{Evolution equation of the spherical harmonic moments}
We follow the same strategy as for the evolution of the background and transform the evolution equation for the distribution function into a coupled set of evolution equations for the spherical harmonic moments
\begin{eqnarray}
\label{eq:deltaCMom}
\delta C_{l,\kt}^{m}(\tau) &=&  \nu_{\rm eff} \int \frac{dp_{\varsigma}}{(2\pi)}  \int \frac{d^2{\bf p}}{(2\pi)^2} \tau^{1/3} p^{\tau} \nonumber \\
&& \qquad Y_{l}^{m}(\phi_{\pt\kt},\thetaP)~\delta f_{\kt}(\tau,\pt,|p_{\varsigma}|) \;, 
\end{eqnarray}
where the azimuthal and polar angles $\phi_{\pt\kt},\thetaP$ are measured with respect to the transverse wave vector $(\kt)$ and the longitudinal rapidity $(\varsigma)$ axis.\footnote{Since physical perturbations of the phase-space distributions $\delta f(\tau, \mathbf{x},\mathbf{p},|p_{\varsigma}|)$ are real-valued, the linearized perturbations in Fourier should also satisfy the condition $\delta f_{-\mathbf k}(\tau,\mathbf{p},|p_{\varsigma}|) =\delta f_{\mathbf k}^{*}(\tau,\mathbf{p},|p_{\varsigma}|)$, such that in terms of the spherical harmonic moments one finds $\delta C^{m}_{\ell,-\kt} (\tau) = (-1)^m  \left( \delta C^{-m}_{\ell,\kt} (\tau)\right)^{*}$ for physical perturbations.} Similarly as for the background, the various components of the energy-momentum tensor are explicitly given in terms of the lowest order ($\ell=0,1,2$) moments, as
\begin{subequations}
\label{eq:EnergyMomentumMomentsAll}
\begin{align}
\tau^{4/3} \delta T^{\tau\tau}_{\kt} &= \sqrt{4\pi} \delta C_{0,\kt}^{0}\;, 
\end{align}
\begin{align}
\delta^{ij} \frac{i\kt^{i}}{|\kt|} \tau^{4/3} \delta T^{\tau j}_{\kt} &= -i\sqrt{\frac{2\pi}{3}} \left( \delta C_{1,\kt}^{+1} - \delta C^{-1}_{1,\kt} \right)\;,  \\
\epsilon^{ij} \frac{i\kt^{i}}{|\kt|} \tau^{4/3} \delta T^{\tau j}_{\kt} &= - ~\sqrt{\frac{2\pi}{3}} \left( \delta C_{1,\kt}^{+1} + \delta C^{-1}_{1,\kt} \right)\;, \\
 \tau^{4/3} (-\tau) \delta T^{\tau \varsigma}_{\kt} &= \sqrt{\frac{4\pi}{3}} \delta C_{1,\kt}^{0}  \;,  
 \end{align}
 \begin{align}
 \delta^{ij} \tau^{4/3}\delta T^{ij}_{\kt} &= \sqrt{\frac{16\pi}{9}} \delta C_{0,\kt}^{0} -  \sqrt{\frac{16\pi}{45}} \delta C_{2,\kt}^{0} \;, \\
\frac{\kt^{i}\kt^{j}}{\kt^2} \tau^{4/3} \delta T^{ij}_{\kt} &=  \sqrt{\frac{4\pi}{9}} \delta C_{0,\kt}^{0} -  \sqrt{\frac{4\pi}{45}} \delta C_{2,\kt}^{0}  \nonumber \\
& + \sqrt{\frac{2\pi}{15}} \left( \delta C_{2,\kt}^{+2} + \delta C_{2,\kt}^{-2} \right) \;,  \\
\epsilon^{lj} \frac{\kt^{i}\kt^{l}}{\kt^2} \tau^{4/3}\delta T^{ij}_{\kt} &=  -i \sqrt{\frac{2 \pi}{15}} \left( \delta C_{2,\kt}^{+2}  - \delta C_{2,\kt}^{-2} \right) \;, 
\end{align}
\begin{align}
\delta^{ij} \frac{i\kt^{i}}{|\kt|} \tau^{4/3} (-\tau) \delta T^{\varsigma j}_{\kt} &=  -i\sqrt{\frac{2 \pi}{15}} \left( \delta C_{2,\kt}^{+1}  - \delta C_{2,\kt}^{-1} \right) \;,  \\
\epsilon^{ij} \frac{i\kt^{i}}{|\kt|} \tau^{4/3} (-\tau) \delta T^{\varsigma j}_{\kt} &= - \sqrt{\frac{2 \pi}{15}} \left( \delta C_{2,\kt}^{+1}  + \delta C_{2,\kt}^{-1} \right)\;, \\
 \tau^{4/3}~\tau^2\delta T^{\varsigma\varsigma}_{\kt} &= \sqrt{\frac{16\pi}{45}}  \delta C_{2,\kt}^{0} + \sqrt{\frac{4\pi}{9}} \delta C_{0,\kt}^{0}\;, 
\end{align}
\end{subequations}
where we have conveniently decomposed all transverse vectors $\delta T^{\tau i},\delta T^{\varsigma i}$ into components parallel, perpendicular and independent w.r.t. to the wave-vector $\kt/|\kt|$ (and similarly for the tensor $\delta T^{ij}$). Due to the residual $\mathbb{Z}_{2}$ symmetry of long. reflection in rapidity, the components $T^{\tau \varsigma}$ and $T^{\varsigma i}$ vanish identically, whereas all other components can in principle be non-zero in our setup. 

Evaluating the various couplings between spherical harmonics using the identities listed in App.~\ref{sec:SphericalHarmonicIdentities}, the evolution equations for the spherical harmonic moments $\delta C_{l}^{m}$ then take the form
\begin{widetext}
\begin{eqnarray}
\label{eq:dC_EOM}
&& \tau \partial_{\tau} \delta C_{l,\kt}^{m} = b_{l,-2}^{m}  \delta C_{l-2,\kt}^{m} + b_{l,0}^{m}  \delta C_{l,\kt}^{m} + b_{l,+2}^{m}  \delta C_{l+2,\kt}^{m} -\frac{i |\kt| \tau }{2} \Big(u_{l,-}^{m} \delta C_{l-1,\kt}^{m+1} + u_{l,+}^{m} \delta C_{l+1,\kt}^{m+1} + d_{l,-}^{m} \delta C_{l-1,\kt}^{m-1} + d_{l,+}^{m} \delta C_{l+1,\kt}^{m-1}\Big) \nonumber \\
&& - \left(\frac{\tau}{\tau_R}\right) \left[ \delta C_{l,\kt}^{m}  + \frac{\delta e_{\kt}}{4e} (C'_{\rm eq})_{l}^{m}  \right] -   \frac{\delta e_{\kt}}{4e} \left(\frac{\tau}{\tau_R}\right) \frac{T(\tau)}{\tau_R} \frac{\partial \tau_R}{\partial T} (C_{\rm eq}-C)_{l}^{m}    \nonumber \\
&& - \left(\frac{\tau}{\tau_R}\right) \frac{\delta u^{\|}_{\kt}}{2}  \Big(u_{l,-}^{m} (C_{\rm eq}-C+C'_{\rm eq})_{l-1}^{m+1} + u_{l,+}^{m} (C_{\rm eq}-C+C'_{\rm eq})_{l+1}^{m+1} + d_{l,-}^{m} (C_{\rm eq}-C+C'_{\rm eq})_{l-1}^{m-1}  + d_{l,+}^{m} (C_{\rm eq}-C+C'_{\rm eq})_{l+1}^{m-1}\Big)   \nonumber \\
&& - \left(\frac{\tau}{\tau_R}\right) \frac{\delta u^{\bot}_{\kt}}{2i}     \Big(u_{l,-}^{m} (C_{\rm eq}-C+C'_{\rm eq})_{l-1}^{m+1} + u_{l,+}^{m} (C_{\rm eq}-C+C'_{\rm eq})_{l+1}^{m+1} - d_{l,-}^{m} (C_{\rm eq}-C+C'_{\rm eq})_{l-1}^{m-1}  - d_{l,+}^{m} (C_{\rm eq}-C+C'_{\rm eq})_{l+1}^{m-1}\Big)   \nonumber \\
\end{eqnarray}
where the coefficients $b_{l}^{m}$ are the same as in Eq.~(\ref{eq:BCoefficientsDef}) and
\begin{eqnarray}
u_{l,-}^{m}&=&+\sqrt{\frac{(l-m)(l-m-1)}{4l^2+1}}\;, \qquad  u_{l,+}^{m}=-\sqrt{\frac{(l+m+1)(l+m+2)}{3+4l(l+2)}}\;, \\
d_{l,-}^{m}&=&-\sqrt{\frac{(l+m)(l+m-1)}{4l^2+1}}\;, \qquad d_{l,+}^{m}=+\sqrt{\frac{(l-m+1)(l-m+2)}{3+4l(l+2)}}\;.
\end{eqnarray}
\end{widetext}
Physically the terms in the first line of Eq.~(\ref{eq:dC_EOM}) correspond to a free-streaming evolution, while the terms in the last few lines capture the relaxation towards equilibrium, including the changes of the equilibrium distribution and background equilibration. By $C'_{\rm eq}$ in Eq.~(\ref{eq:dC_EOM}) we denote the moments
\begin{eqnarray}
(C'_{\rm eq}) _{l}^{m}(\tau)&=& \int \frac{dp_{\varsigma}}{(2\pi)}  \int \frac{d^2{\bf p}}{(2\pi)^2} \tau^{1/3} p^{\tau}   \\
&& \qquad Y_{l}^{m}(\phi_{\pt\kt},\thetaP)~\left( \frac{p^{\tau}}{T(\tau)} \right) f_{\rm eq}'\left( \frac{p^{\tau}}{T(\tau)} \right)\;, \nonumber
\end{eqnarray}
which can be determined via integration by parts as
\begin{eqnarray}
(C'_{\rm eq}) _{l}^{m}(\tau)&=& -4 (C_{\rm eq})_{l}^{m}\;,
\end{eqnarray}
with the equilibrium moments $(C_{\rm eq})_{l}^{m}$ determined by Eq.~(\ref{eq:CeqVals}). By inserting the relations in Eqns.~(\ref{eq:EnergyMomentumMomentsAll}) into the linearized Landau matching conditions in Eq.~(\ref{eq:linLandmatch}), one also finds that the linearized perturbations of the energy density $\delta e_{\kt}$ and long. and transverse flow velocities $\delta u^{\|}_{\kt}$ and $\delta u^{\bot}_{\kt}$  are given by
\begin{eqnarray}
 \tau^{4/3} \delta e_{\kt}(\tau)&=&\sqrt{4\pi} \delta C_{0,{\bf k}}^{0}(\tau)\;,  \\
\tau^{4/3} (e+p_T)  \delta u_{\bf k}^{\|}(\tau) &=& -\sqrt{\frac{2\pi}{3}}\Big(\delta C_{1,{\bf k}}^{1}(\tau) - \delta C_{1,{\bf k}}^{-1}(\tau) \Big)  \;, \nonumber \\
\tau^{4/3} (e+p_T)  \delta u_{\bf k}^{\bot}(\tau)&=& ~i\sqrt{\frac{2\pi}{3}} \Big(\delta C_{1,{\bf k}}^{1}(\tau) + \delta C_{1,{\bf k}}^{-1}(\tau)\Big) \;,  \nonumber
\end{eqnarray}
such that the evolution equations for the moments in Eq.~(\ref{eq:dC_EOM}) form a closed set of equations. We further note that by virtue of the decomposition into spherical harmonic moments, $Y_{l}^{m}(\phi_{\pt\kt},\thetaP)$, the information on the direction of the transverse wave-vector $\kt$ has disappeared from the evolution equation, which as a consequence of the azimuthal rotation symmetry of the background only depends on the magnitude of the wave-vector $|\kt|$. 

Before we address the physically relevant initial conditions for the moments $\delta C_{l,\kt}^{m}$, we also note that it is often useful to consider the evolution equation at a fixed value of propagation phase 
\begin{eqnarray}
\kappa=|\kt| (\tau-\tau_0)
\end{eqnarray}
rather than a fixed value for the wave-number $|\kt|$. By changing the variables from $|\kt|$ to $\kappa=|\kt|(\tau-\tau_0)$ for the moments, the time derivate needs to be evaluated according to
\begin{eqnarray}
\left. \tau \partial_{\tau}\right|_{\kt} =  \left. \tau \partial_{\tau}\right|_{\kt (\tau-\tau_0)} + \frac{\tau}{\tau-\tau_0} |\kt| (\tau-\tau_0) \left.  \partial_{|\kt|(\tau-\tau_0)} \right|_{\tau}\;, \nonumber \\
\end{eqnarray}
such that the evolution equation for the moments receives one additional term associated with this change of variables. Similarly, for a conformal system, it is also convenient to express the evolution in terms of the scaled evolution time $x=\tau/\tau_R$ introduced in Eq.~(\ref{eq:xEvoDef}). Starting from Eq.~(\ref{eq:dC_EOM}) and denoting
$s(\tau)=\frac{\tau-\tau_0}{\tau}$ , with
\begin{eqnarray}
\label{eq:SFacEvo}
a(x) x \partial_{x} s(x) =(1-s(x))\;,
\end{eqnarray}
to perform these changes, the equation of motion for the moments then takes the form
\begin{widetext}
\begin{eqnarray}
\label{eq:dC_EOM_Scaling}
&& \Big[  s(x) a(x) x\partial_{x} +\kappa \partial_{\kappa} \Big]  \delta C_{l,\kappa}^{m} =  \nonumber \\
&&~~~~ s(x) \Big(b_{l,-2}^{m}  \delta C_{l-2,\kappa}^{m} + b_{l,0}^{m}  \delta C_{l,\kappa}^{m} + b_{l,+2}^{m}   \delta C_{l+2,\kappa}^{m}\Big) -\frac{i \kappa }{2} \Big(u_{l,-}^{m} \delta C_{l-1,\kappa}^{m+1} + u_{l,+}^{m} \delta C_{l+1,\kappa}^{m+1} + d_{l,-}^{m} \delta C_{l-1,\kappa}^{m-1} + d_{l,+}^{m} \delta C_{l+1,\kappa}^{m-1}\Big) \nonumber \\
&&  - x s(x)  \left[ \delta C_{l,\kappa}^{m}  + \frac{\delta e_{\kappa}}{4e} (C'_{\rm eq})_{l}^{m}  \right] -  x s(x)  \frac{\delta e_{\kappa}}{4e} \frac{T(\tau)}{\tau_R} \frac{\partial \tau_R}{\partial T} (C_{\rm eq}-C)_{l}^{m}    \nonumber \\
&& - x s(x)  \frac{\delta u^{\|}_{\kappa}}{2}  \Big(u_{l,-}^{m} (C_{\rm eq}-C+C'_{\rm eq})_{l-1}^{m+1} + u_{l,+}^{m} (C_{\rm eq}-C+C'_{\rm eq})_{l+1}^{m+1} + d_{l,-}^{m} (C_{\rm eq}-C+C'_{\rm eq})_{l-1}^{m-1}  + d_{l,+}^{m} (C_{\rm eq}-C+C'_{\rm eq})_{l+1}^{m-1}\Big)   \nonumber \\
&& - x s(x)  \frac{\delta u^{\bot}_{\kappa}}{2i}  \Big(u_{l,-}^{m} (C_{\rm eq}-C+C'_{\rm eq})_{l-1}^{m+1} + u_{l,+}^{m} (C_{\rm eq}-C+C'_{\rm eq})_{l+1}^{m+1} - d_{l,-}^{m} (C_{\rm eq}-C+C'_{\rm eq})_{l-1}^{m-1}  - d_{l,+}^{m} (C_{\rm eq}-C+C'_{\rm eq})_{l+1}^{m-1}\Big)   \nonumber \\
\end{eqnarray}
\end{widetext}
which we will employ below to obtain the numerical solution for the evolution of perturbations. 

Based on Eq.~(\ref{eq:dC_EOM_Scaling}), one also explicitly observes that  in the limit $\tau_0/\tau_R \to 0$ where $s(x)=1$ is a fixed point of Eq.~(\ref{eq:SFacEvo}), the solution to the evolution equation for the moments $\delta C_{l,\kt}^{m}$ only depends on the propagation phase $\kappa=|\kt|(\tau-\tau_0)$ and the scaled evolution time $x=\tau/\tau_R = \frac{T(\tau)\tau}{5 \eta/s}$. While this conformal scaling behavior was empirically observed in \cite{Kurkela:2018vqr,Kurkela:2018wud} from numerical solutions of the Boltzmann equation in QCD kinetic theory at different values of the coupling strength $\lambda=g^2 N_c$, it is interesting to point out that in the present context the conformal scaling behavior directly manifests itself at the level of the equations of motion.

\subsection{Initial conditions for energy and momentum perturbations}
\label{sec:InitialConditons}
So far we have discussed, the evolution equations for linearized perturbations on top of a Bjorken background. Now in order to apply this framework to describe the early time dynamics of high-energy heavy-ion collisions, the equations of motion need to be supplemented by suitable initial conditions, which describe the associated change of the phase-space density at initial time. While in principle one could imagine a large variety of different initial conditions, we will follow  \cite{Kurkela:2018vqr,Kurkela:2018wud} and only consider the response of the system to changes of the conserved quantities of the system, associated with initial energy and momentum perturbations as detailed below.
\subsubsection{Energy perturbations}
We follow the arguments of \cite{Kurkela:2018vqr} and associate initial energy perturbations with an infinitesimal change of the energy scale of the background distribution, such that the associated phase-space distribution for energy perturbations is given by
\begin{eqnarray}
\delta f_{\kt}(\tau_0,\pt,|p_{\varsigma}|)= -\left(\frac{|\pt|}{3} \partial_{|\pt|} f_{BG}^{(0)}(|\pt|,p_{\varsigma})\right) e^{-i\kt\frac{\pt}{|\pt|}\tau_0}\;, \nonumber \\
\end{eqnarray}
where we introduced the phase-factor $e^{-i\kt\frac{\pt}{|\pt|}\tau_0}$ to account for the free-streaming evolution at early times $\tau < \tau_0 \ll \tau_R$. Based on the explicit form of the initial background distribution in Eq.~(\ref{eq:inidistfunc}), the integrals for the moments $\delta C_{l}^{m}$ in Eq.~(\ref{eq:deltaCMom}) can be evaluated using
\begin{eqnarray}
\frac{1}{2\pi} \int_{0}^{2\pi} d\phi_{\pt\kt} e^{-i|\kt|\tau_0 \cos(\phi_{\pt\kt})}~e^{im\phi_{\pt\kt}} = (-i)^{m} J_{m}(|\kt|\tau_0)\;, \nonumber \\
\end{eqnarray}
along with Eq.~(\ref{eq:CMomInitialCond}) yielding
\begin{eqnarray}
\delta C_{l,\kt}^{m}(\tau_0) = \tau_0^{1/3} (e\tau)_0 (-i)^{m} J_{m}(|\kt|\tau_0) y_{l}^{m} P_{l}^{m}(0)\;, \nonumber \\
\end{eqnarray}
where $(e\tau)_0$ denotes the asymptotic energy density of the background (c.f. Sec.~\ref{sec:back}). Specifically, for the energy and momentum density one has
\begin{eqnarray}
\frac{\delta e_{\kt}(\tau_0)}{e} &=&  ~~~J_{0}(|\kt|\tau)\;, \\
\frac{(e+p_T)}{e} \delta u^{\|}_{\kt}(\tau_0) &=& -i  J_{1}(|\kt|\tau)\;, \\
\frac{(e+p_T)}{e} \delta u^{\bot}_{\kt}(\tau_0) &=& 0\;,
\end{eqnarray}
reproducing the result of \cite{Kurkela:2018vqr} for the free-streaming response function.
\subsubsection{Momentum perturbations}
Similarly, we associate initial momentum perturbations with an infinitesimal change of the transverse velocity of the background,  such that following the arguments of \cite{Kurkela:2018vqr} the associated phase-space distribution for momentum perturbations is given by
\begin{eqnarray}
\delta f_{\kt}^{i}(\tau_0,\pt,p_{\varsigma})= 2\left(\frac{\pt^{i}}{3}\partial_{|\pt|} f_{BG}^{(0)}(|\pt|,p_{\varsigma})\right) e^{-i\kt\frac{\pt}{|\pt|}\tau_0}\;, \nonumber \\
\end{eqnarray}
where the index $i$ contains the information about the direction of the initial momentum perturbation. Decomposing $\delta f_{\kt}^{i}$ into the directions parallel and perpendicular to the wave-vector, we can distinguish between longitudinal and transverse momentum perturbations
\begin{eqnarray}
&&\delta f_{\kt}^{\|}(\tau_0,\pt,p_{\varsigma})= \\
&& \qquad 2\cos(\phi_{\pt\kt})  \left(\frac{|\pt|}{3} \partial_{|\pt|} f_{BG}^{(0)}(|\pt|,p_{\varsigma})\right) e^{-i\kt\frac{\pt}{|\pt|}\tau_0}\;, \nonumber \\
&&\delta f_{\kt}^{\bot}(\tau_0,\pt,p_{\varsigma})= \\
&& \qquad 2\sin(\phi_{\pt\kt})  \left(\frac{|\pt|}{3} \partial_{|\pt|} f_{BG}^{(0)}(|\pt|,p_{\varsigma})\right) e^{-i\kt\frac{\pt}{|\pt|}\tau_0}\;. \nonumber
\end{eqnarray}
Evaluating the moments, one then finds
\begin{eqnarray}
&& \delta C_{l,\kt}^{m \|}(\tau_0) = \\ 
&& -i \tau_0^{1/3} (e\tau)_0 (-i)^{m} \Big(J_{m+1}(|\kt|\tau_{0}) - J_{m-1}(|\kt|\tau_{0}) \Big) y_{l}^{m} P_{l}^{m}(0)\;, \nonumber
\end{eqnarray}
for longitudinal momentum perturbations and similarly for transverse momentum perturbations
\begin{eqnarray}
&& \delta C_{l,\kt}^{m \bot}(\tau_0) = \\ 
&& - \tau_0^{1/3} (e\tau)_0 (-i)^{m} \Big(J_{m+1}(|\kt|\tau_{0}) + J_{m-1}(|\kt|\tau_{0}) \Big) y_{l}^{m} P_{l}^{m}(0)\;. \nonumber
\end{eqnarray}
Specifically, for the energy and momentum density one has
\begin{eqnarray}
\frac{\delta e_{\kt}(\tau_0)}{e} &=&  -2i J_{1}(|\kt|\tau_{0})\;, \\
\frac{(e+p_T)}{e} \delta u^{\|}_{\kt}(\tau_0) &=& 2\frac{J_{1}(|\kt| \tau)}{|\kt| \tau_{0}} - 2 J_{2}(|\kt|\tau_{0})\;, \\
\frac{(e+p_T)}{e} \delta u^{\bot}_{\kt}(\tau_0) &=& 0\;,
\end{eqnarray}
for longitudinal momentum perturbations, whereas for transverse momentum perturbations
\begin{eqnarray}
\frac{\delta e_{\kt}(\tau_0)}{e} &=&  0\;, \\
\frac{(e+p_T)}{e} \delta u^{\|}_{\kt}(\tau_0) &=& 0\;, \\
\frac{(e+p_T)}{e} \delta u^{\bot}_{\kt}(\tau_0) &=& 2\frac{J_{1}(|\kt| \tau)}{|\kt| \tau} \;,
\end{eqnarray}
in agreement with  \cite{Kurkela:2018vqr}.

\begin{figure*}[t!]
\begin{center}
\begin{minipage}{0.49\textwidth}
\includegraphics[width=\textwidth]{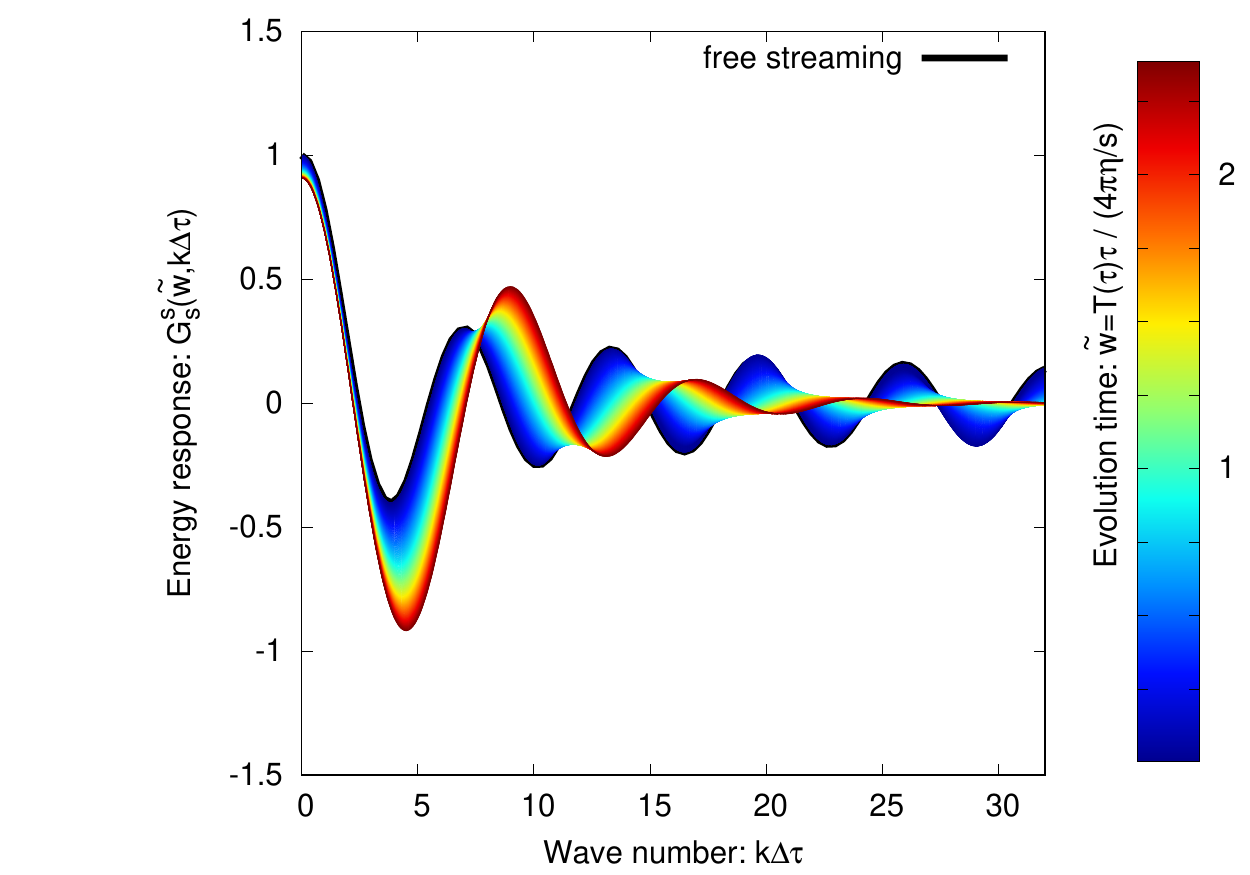}
\includegraphics[width=\textwidth]{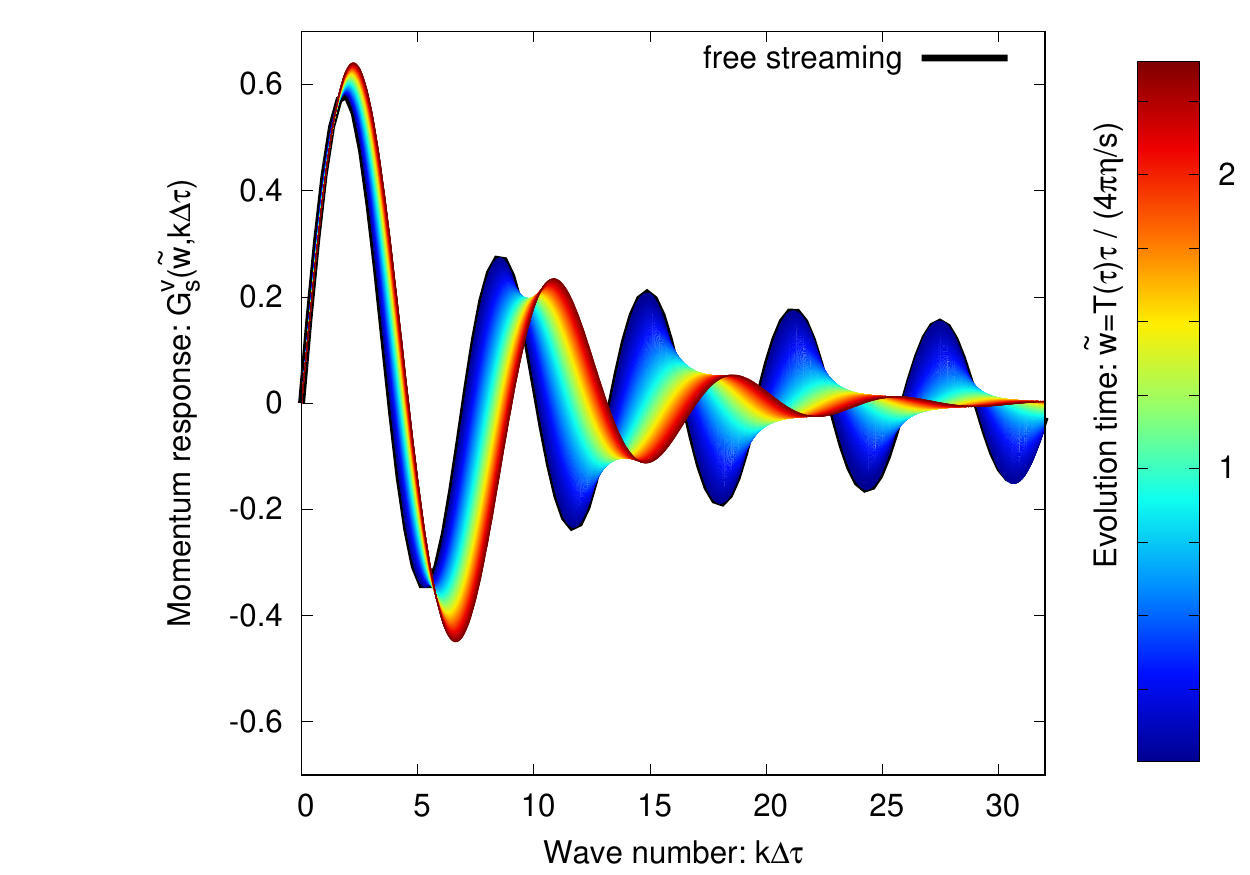}
\end{minipage}
\begin{minipage}{0.49\textwidth}
\includegraphics[width=\textwidth]{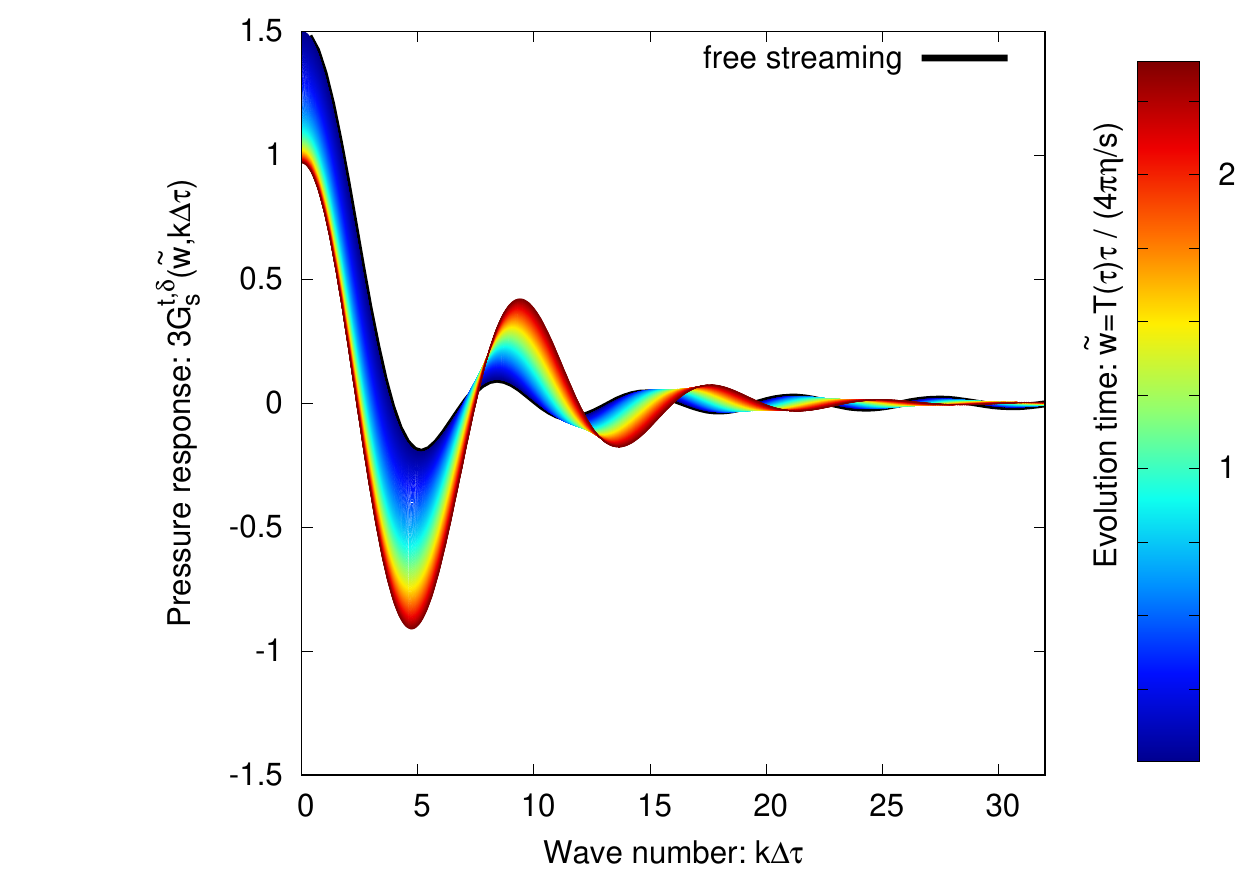}
\includegraphics[width=\textwidth]{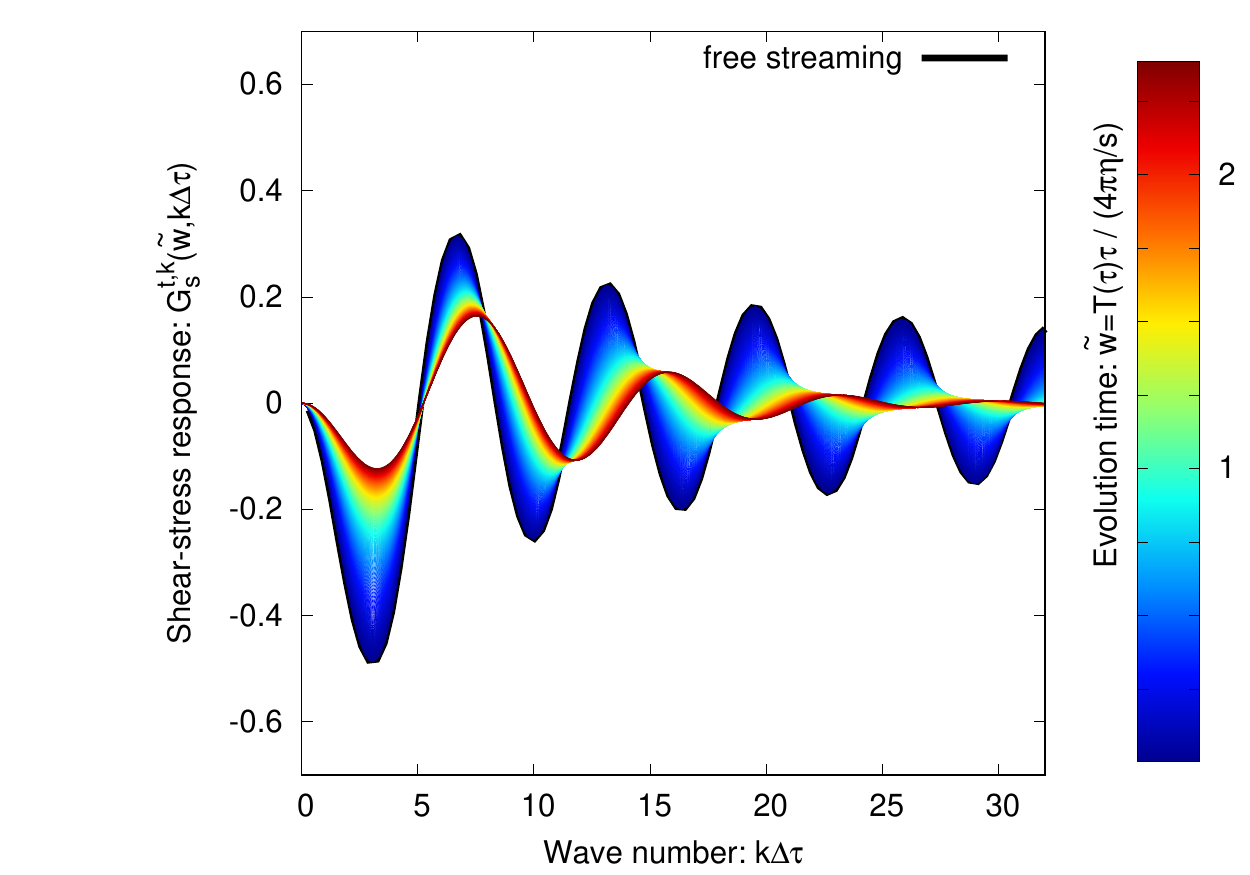}
\end{minipage}
\caption{\label{fig:EnergyResponseK} Evolution of spectrum of energy-momentum perturbations in response to an initial energy perturbation. Different curves in each panel correspond to different evolution times $T(\tau)\tau/(4\pi \eta/s)$; different panels show the response of the different components of the energy-momentum tensor as a function of the wave-number $|\kt|(\tau-\tau_0)$.}
\end{center}
\end{figure*}

\begin{figure*}[t!]
\begin{center}
\begin{minipage}{0.49\textwidth}
\includegraphics[width=\textwidth]{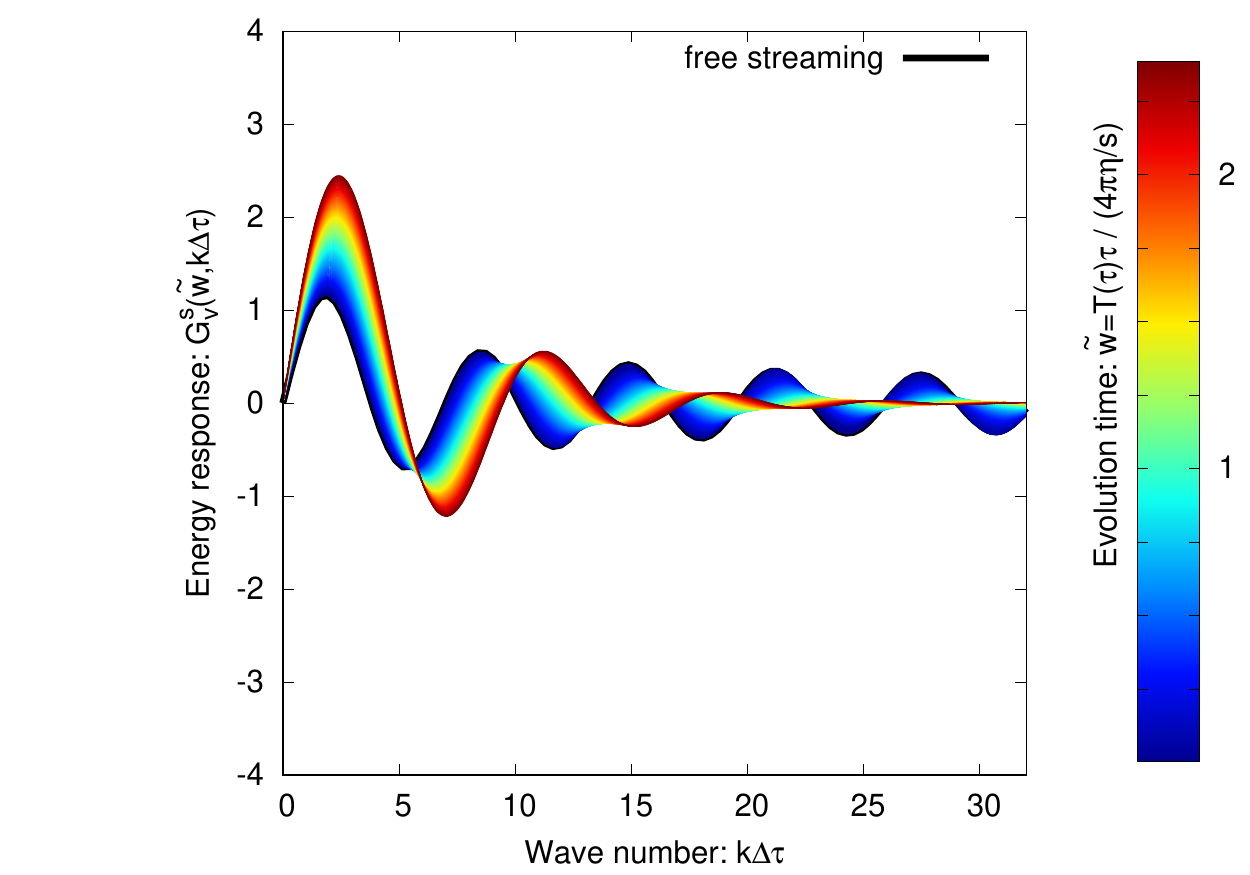}
\includegraphics[width=\textwidth]{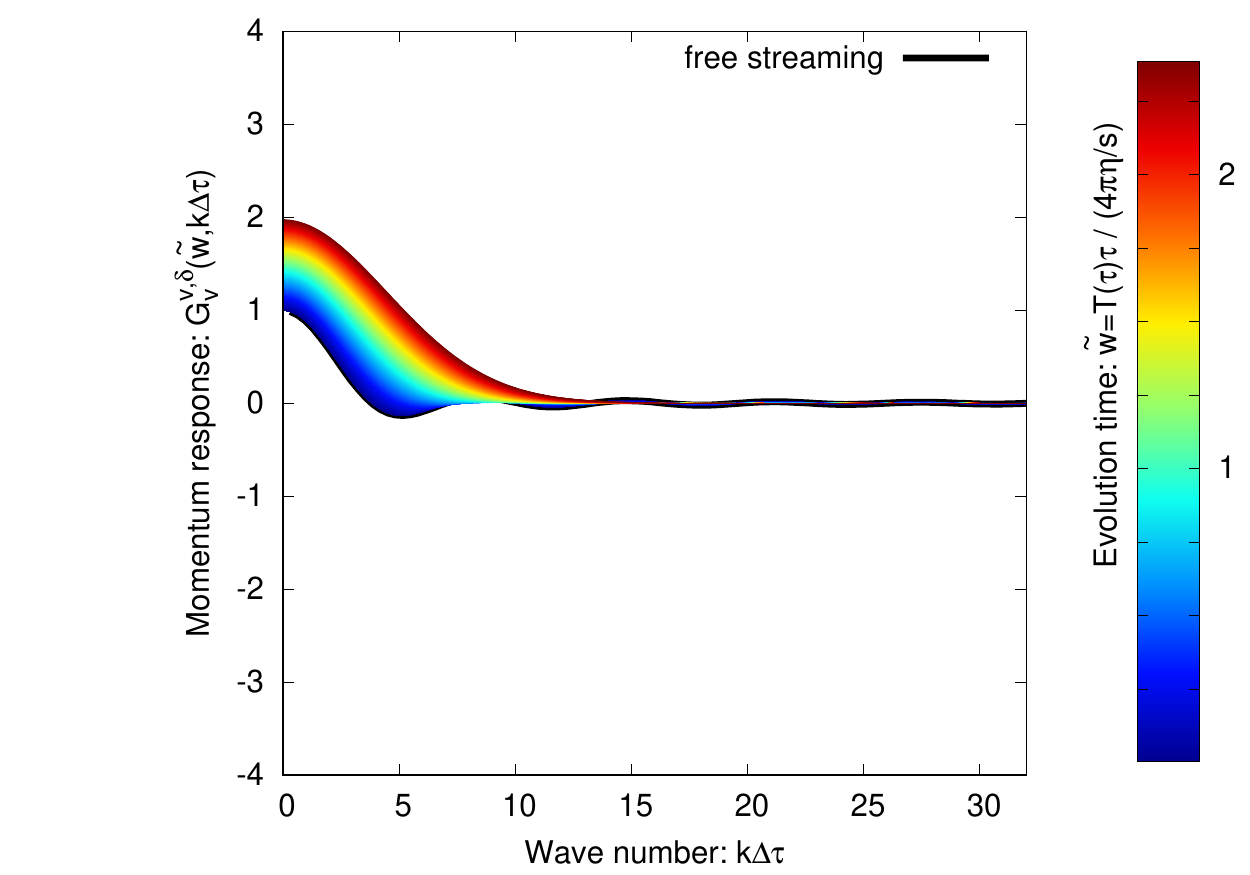}
\includegraphics[width=\textwidth]{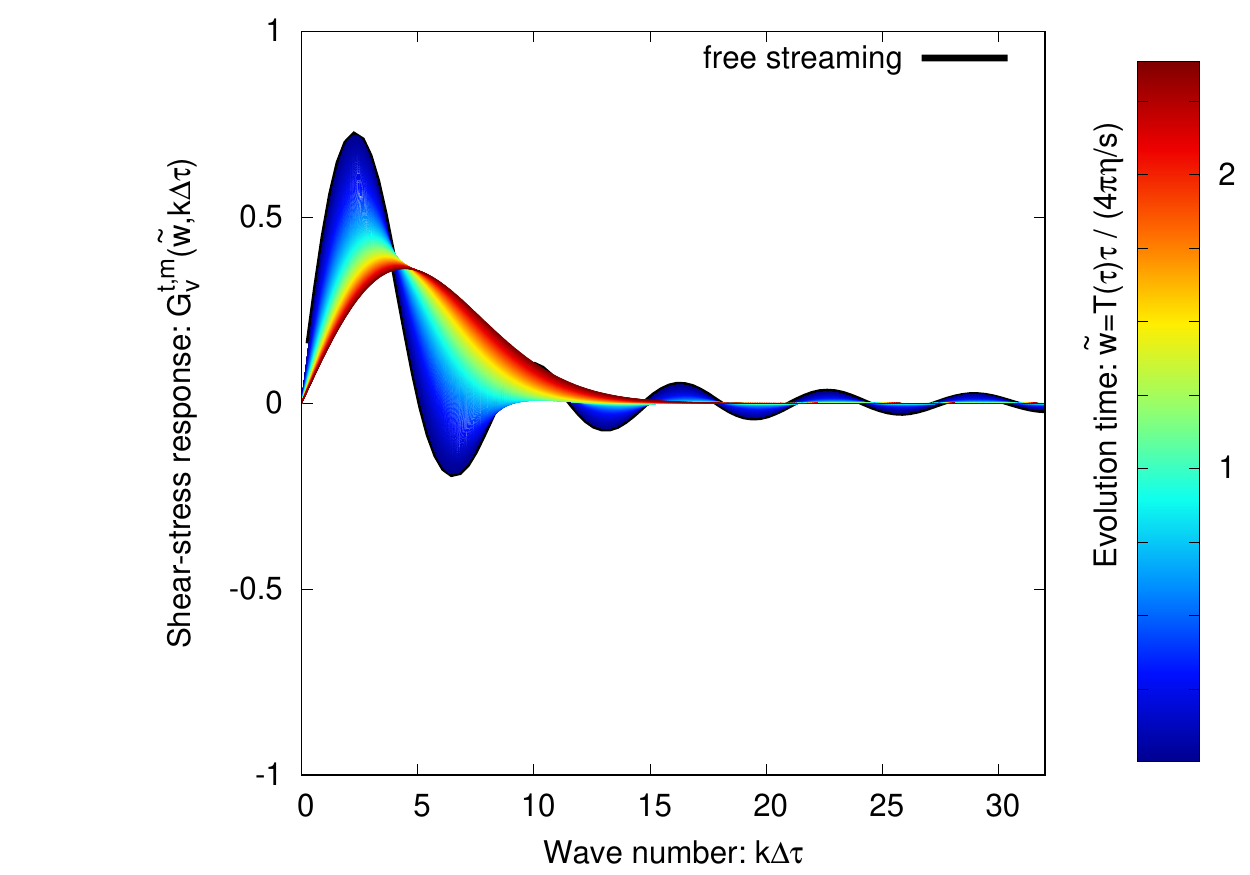}
\end{minipage}
\begin{minipage}{0.49\textwidth}
\includegraphics[width=\textwidth]{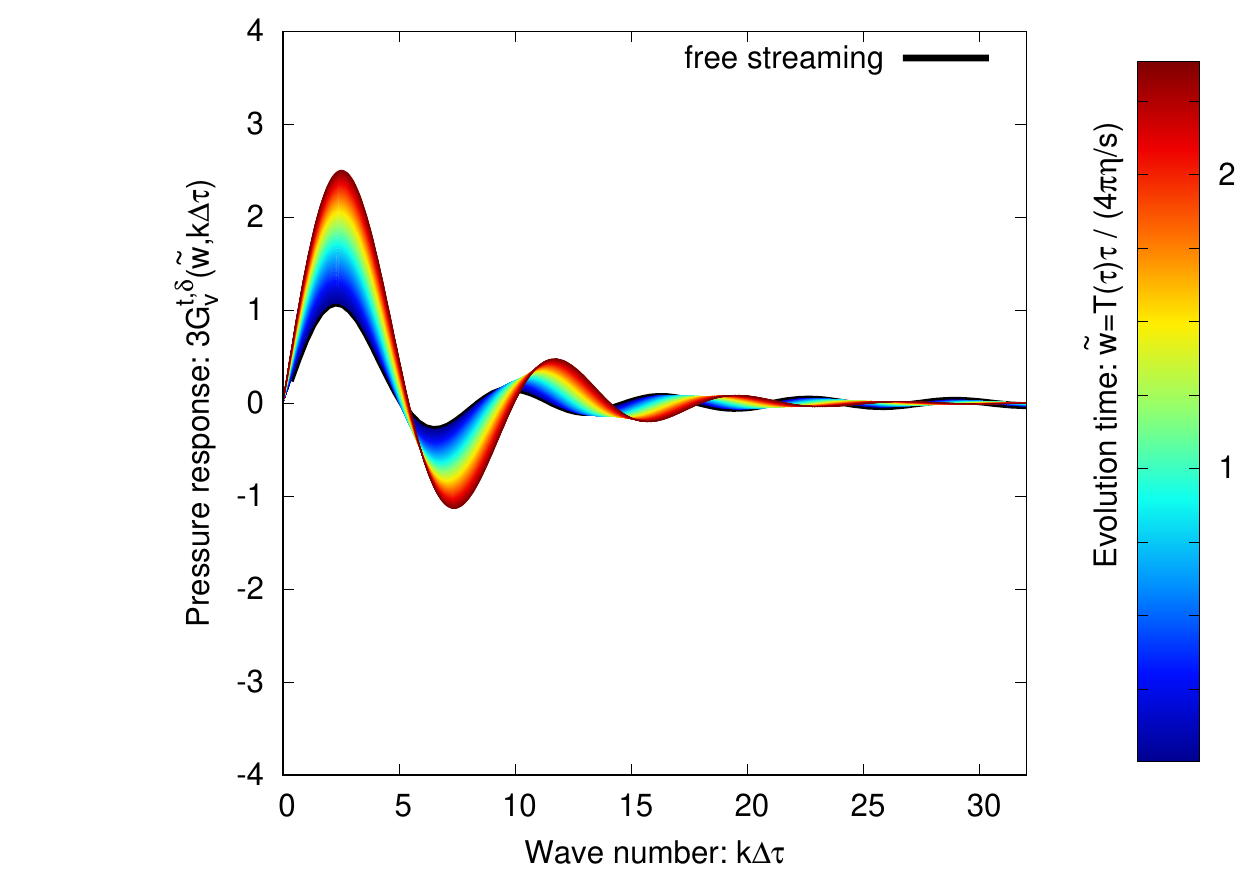}
\includegraphics[width=\textwidth]{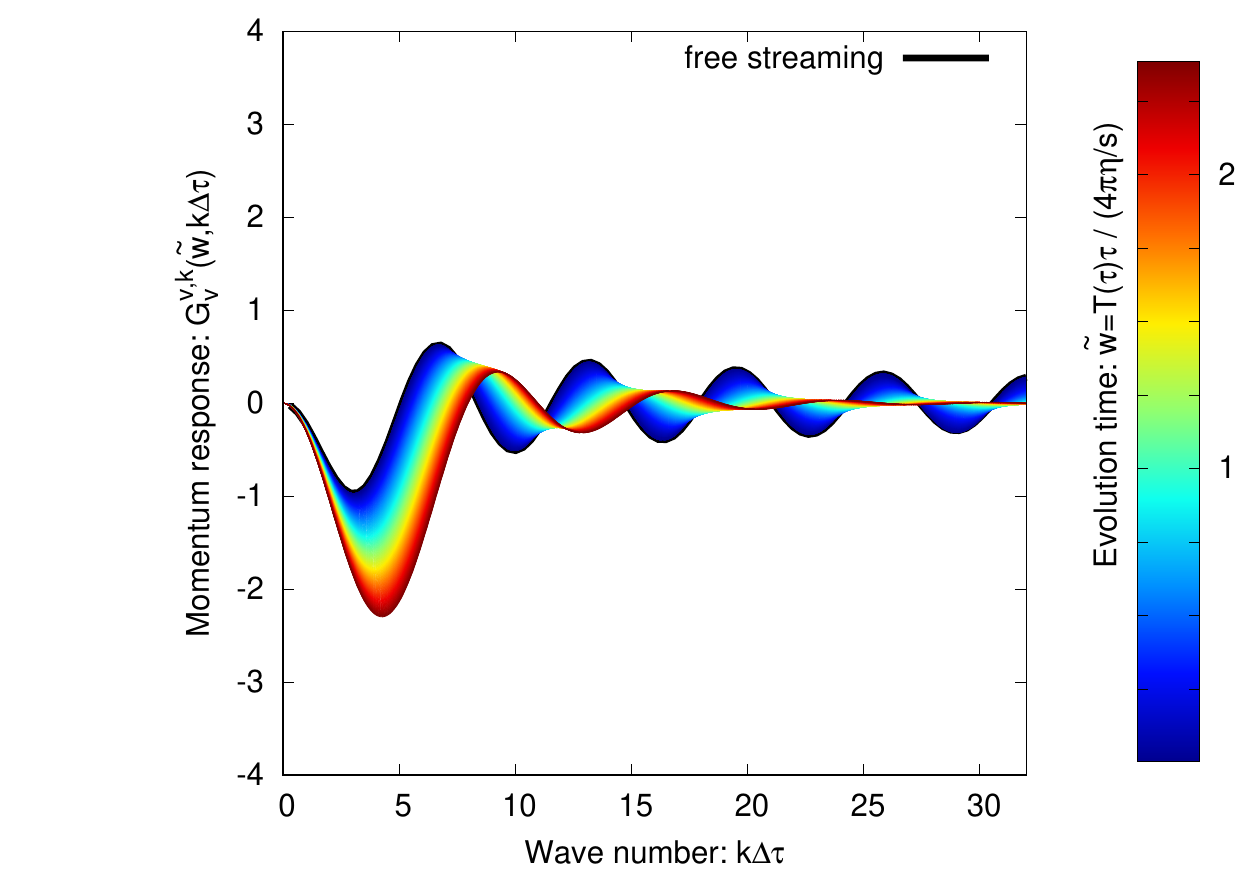}
\includegraphics[width=\textwidth]{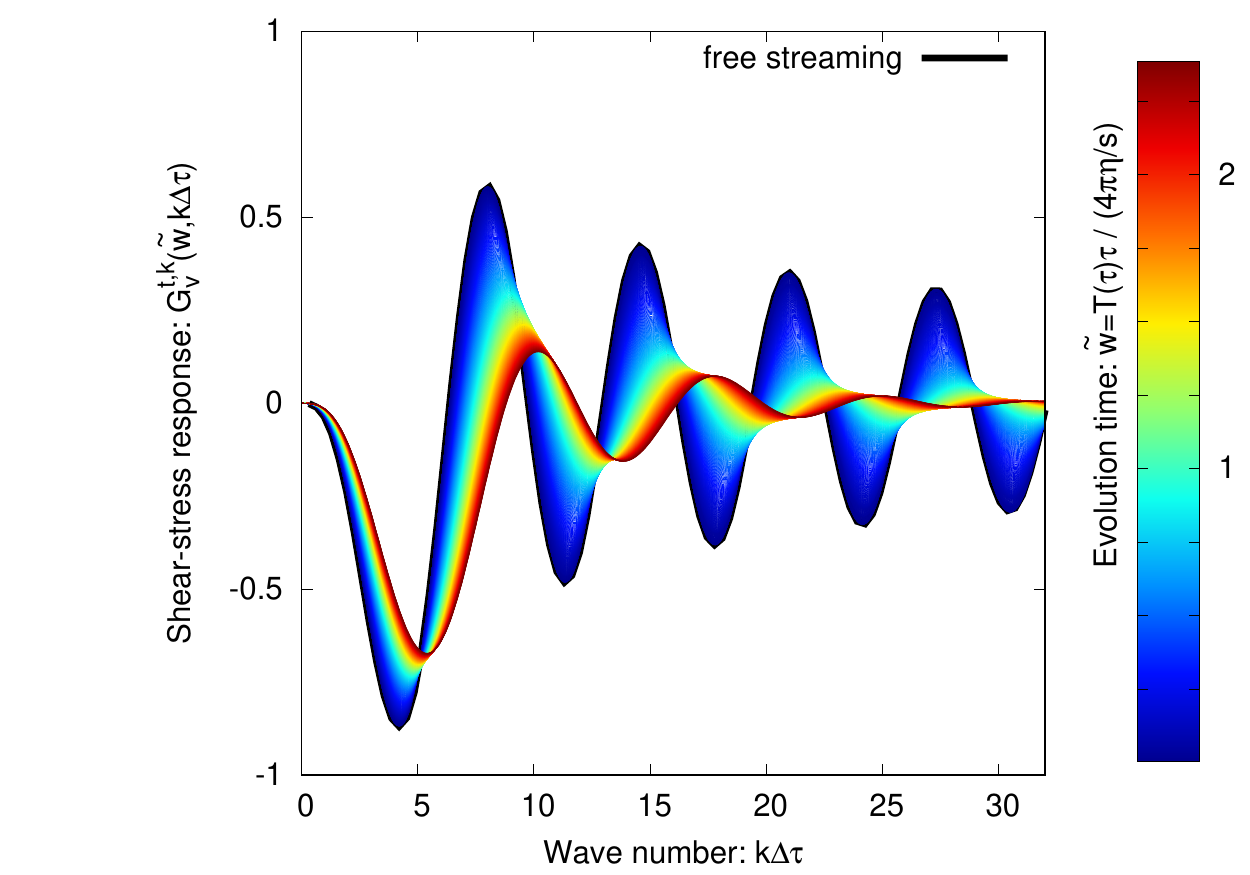}
\end{minipage}
\caption{\label{fig:MomentumResponseK} Evolution of spectrum of energy-momentum perturbations in response to an momentum energy perturbation. Different curves in each panel correspond to different evolution times $T(\tau)\tau/(4\pi \eta/s)$; different panels show the response of the different components of the energy-momentum tensor as a function of the wave-number $|\kt|(\tau-\tau_0)$.}
\end{center}
\end{figure*}

\section{Non-equilibrium Green's functions of the energy-momentum tensor}
\label{sec:GreensFunctions}

We now proceed to the calculation of the response of the energy-moment tensor to initial energy and momentum perturbations, based on numerical solutions of the evolution equations for the moments $\delta C_{l,\kt}^{m}$ starting from the initial condition described in the previous section. Since the set of moments $\delta C_{l,\kt}^{m}$ contain an overwhelming amount of information, we will therefore restrict our attention to the evolution of the low order moments $\delta C_{l,\kt}^{m}$ with $\ell \leq 2$, which can be related to the various components of the energy momentum tensor $\delta T^{\mu\nu}_{\kt}$ according to Eq.~(\ref{eq:EnergyMomentumMomentsAll}). Instead of investigating the dynamics of individual moments $\delta C_{l,\kt}^{m}$ directly, we find it more insightful to consider the linear response functions $G^{\mu\nu}_{\alpha\beta}(\kt,\tau,\tau_0)$ introduced in \cite{Kurkela:2018vqr,Kurkela:2018wud}, such that
\begin{eqnarray}
\label{eq:GreensFunctionDef}
\frac{\delta T^{\mu\nu}_{\kt}(\tau)}{e(\tau)}=\frac{1}{2}~G^{\mu\nu}_{\alpha\beta}(\kt,\tau,\tau_0) \frac{\delta T^{\alpha\beta}_{\kt}(\tau_0)}{e(\tau_0)}\;.
\end{eqnarray}
Noteably, the Green's functions $G^{\mu\nu}_{\alpha\beta}(\kt,\tau,\tau_0)$ provide the builiding block of the pre-equilibrium computer code \kompost~\cite{KoMPoSTGit} and can be obtained in terms of linear combinations of the moments as described below.  Since we are primarily interested in the limit $\tau_0/\tau_{R} \to 0$, where the kinetic theory framework describes the equilibration process of the system all the way from very early times up to the onset of hydrodynamic behavior, we will drop the explicit dependence on $\tau_0$ in the following to lighten the notation. By expressing the response to an \emph{initial energy perturbation} in the following basis of scalars (s), vectors (v) and tensors (t)
\begin{subequations}
\label{eq:GsKBasis}
\begin{align}
G^{\tau\tau}_{\tau\tau}(\kt,\tau)= G_{s}^{s}(\kappa,x)\;, 
\end{align}
\begin{align}
G^{\tau i}_{\tau\tau}(\kt,\tau) = -i \frac{\kt^{i}}{|\kt|} G_{s}^{v}(\kappa,x)\;, 
\end{align}
\begin{align}
G^{ij}_{\tau\tau}(\kt,\tau) =  \delta^{ij} G_{s}^{t,\delta}(\kappa,x)  +  \frac{\kt^{i}\kt^{j}}{\kt^2} G_{s}^{t,k}(\kappa,x)\;,
\end{align}
\end{subequations}
and adapting the normalization $\delta e(\tau_0)/e(\tau_0)=1$, the relevant response functions can then be determined from (c.f. Appendix B of \cite{Kurkela:2018vqr})
\begin{subequations}
\begin{align}
G_{s}^{s}(\kappa,x)&=\frac{\delta e_{\kappa}(x)}{e(x)}\;, 
\end{align}
\begin{align}
G_{s}^{v}(\kappa,x) &= \delta^{ij} \frac{i \kt^{i}}{|\kt|} \frac{\delta T^{0j}_{\kappa}(x)}{e(x)}\;,  
\end{align}
\begin{align}
G_{s}^{t,\delta}(\kappa,x) &= \left[\delta^{ij} -\frac{\kt^{i}\kt^{j}}{\kt^2} \right] \frac{\delta T^{ij}_{\kappa}(x)}{e(x)}\;, \\
G_{s}^{t,k}(\kappa,x) &= \left[ 2\frac{\kt^{i}\kt^{j}}{\kt^2} - \delta^{ij}  \right] \frac{\delta T^{ij}_{\kappa}(x)}{e(x)}\;. \nonumber
\end{align}
\end{subequations}
Similarly, the response to an \emph{initial momentum perturbation} can be characterized by a set of six independent response functions, 
\begin{subequations}
\label{eq:GvKBasis}
\begin{align}
G^{\tau\tau}_{\tau l}(\kt,\tau) &= -i \frac{\kt^{l}}{|\kt|} G_{v}^{s}(\kappa,x)\;, 
\end{align}
\begin{align}
G^{\tau i}_{\tau l}(\kt,\tau) &=\delta^{ij} G_{v}^{\delta}(\kappa,x)  +  \frac{\kt^{i}\kt^{j}}{\kt^2} G_{v}^{k}(\kappa,x)\;,
\end{align}
\begin{align}
G^{ij}_{\tau l}(\kt,\tau) &=  -i \frac{\kt^{l}}{|\kt|}  \delta^{ij} G_{v}^{t,\delta}(\kappa,x) -i \frac{\kt^{i}\kt^{j} \kt^{k}}{|\kt|^3} G_{v}^{t,k}(\kappa,x) \nonumber \\
& - i \frac{\delta^{il}\kt^{j} + \delta^{jl} \kt^{i}}{2|\kt|}  G_{v}^{t,m}(\kappa,x)
\end{align}
\end{subequations}
which upon adapting the normalization condition $\delta T^{\tau i}_{\tau j,\kappa}/e(\tau_0)=\delta^{ij}$, where as in Eq.~(\ref{eq:GreensFunctionDef}) upper indices ($\tau i$)  refer to the respective components of the energy-momentum tensor and lower indices ($\tau j$) indicate the direction of the initial momentum perturbation, are given by (c.f. Appendix B of \cite{Kurkela:2018vqr})
\begin{subequations}
\begin{align}
G_{v}^{s}(\kappa,x) &=\delta^{ij} \frac{i \kt^{i}}{|\kt|} \frac{\delta T^{\tau\tau}_{\tau j,\kappa}(x)}{\delta e(x)} \;, 
\end{align}
\begin{align}
G_{v}^{v,\delta}(\kappa,x)  &= \left[\delta^{ij} -\frac{\kt^{i}\kt^{j}}{\kt^2} \right]  \frac{\delta T^{\tau i}_{\tau j,\kappa}(x)}{\delta e(x)} \;,  \\
G_{v}^{v,k}(\kappa,x)  &= \left[ 2\frac{\kt^{i}\kt^{j}}{\kt^2} - \delta^{ij}  \right]  \frac{\delta T^{\tau i}_{\tau j,\kappa}(x)}{\delta e(x)} \;,  
\end{align}
\begin{align}
&G_{v}^{t,\delta}(\kappa,x)  = \left[\delta^{ij} -\frac{\kt^{i}\kt^{j}}{\kt^2} \right] \frac{i \kt^{l}}{|\kt|} \frac{\delta T^{ij}_{\tau l,\kappa}(x)}{\delta e(x)} \\
&G_{v}^{t,m}(\kappa,x)  = 2 \frac{i \kt^{i}}{|\kt|} \left[\delta^{jl} -\frac{\kt^{j}\kt^{l}}{\kt^2} \right] \frac{\delta T^{ij}_{\tau l,\kappa}(x)}{\delta e(x)} \\
&G_{v}^{t,k} (\kappa,x) = \\
&\left( \left[2\frac{\kt^{i}\kt^{j}}{\kt^2} - \delta^{ij}  \right] \frac{i \kt^{l}}{|\kt|} - 2 \frac{i \kt^{i}}{|\kt|} \left[\delta^{jl} -\frac{\kt^{j}\kt^{l}}{\kt^2} \right] \right) \frac{\delta T^{ij}_{\tau l,\kappa}(x)}{\delta e(x)}\;.  \nonumber
\end{align}
\end{subequations}
While Eqns.~(\ref{eq:GsKBasis}) and (\ref{eq:GvKBasis})  provide a basis for expressing the response of the energy-momentum tensor $G^{\mu\nu}_{\alpha\beta}(\kt,\tau,\tau_0)$ in wave-number ($\kt$) space, we will also be interested in the response of the energy-momentum tensor in coordinate space $G^{\mu\nu}_{\alpha\beta}(\xt-\xt_0,\tau,\tau_0)$, where an analogous decomposition can be performed w.r.t. to the vector $\xt-\xt_0$. We will refrain from presenting all the details and instead refer the interested reader to Appendix B of \cite{Kurkela:2018vqr}.

\begin{figure*}[t!]
\begin{center}
\begin{minipage}{0.49\textwidth}
\includegraphics[width=\textwidth]{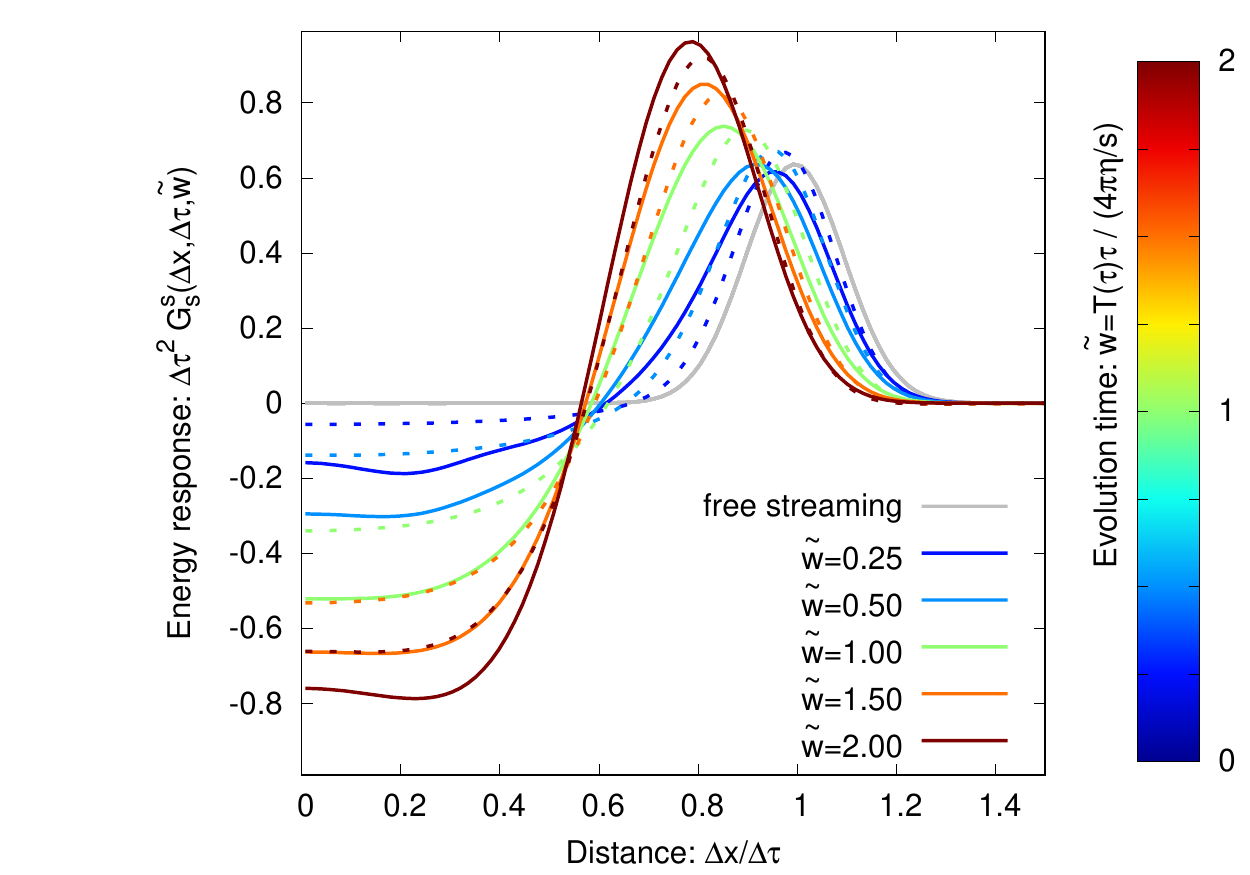}
\includegraphics[width=\textwidth]{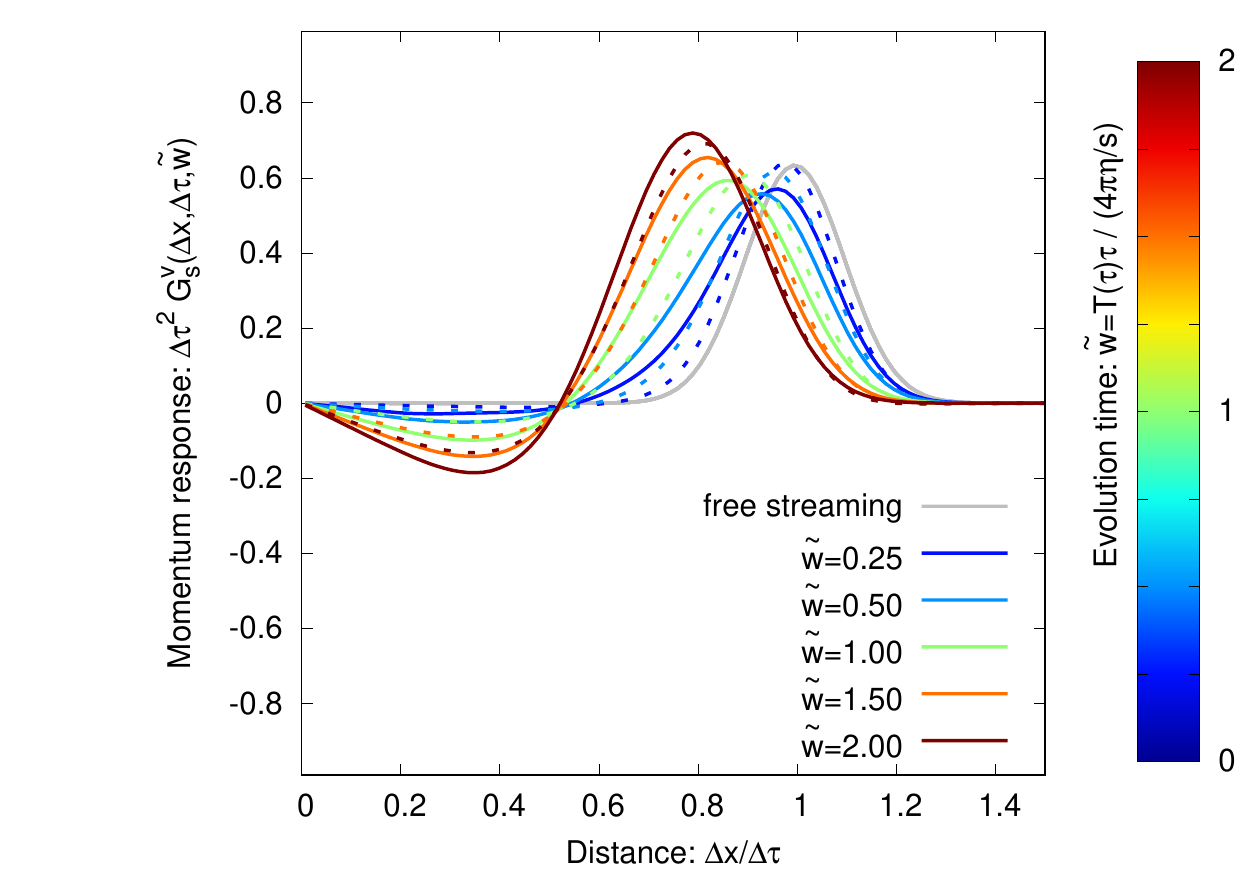}
\end{minipage}
\begin{minipage}{0.49\textwidth}
\includegraphics[width=\textwidth]{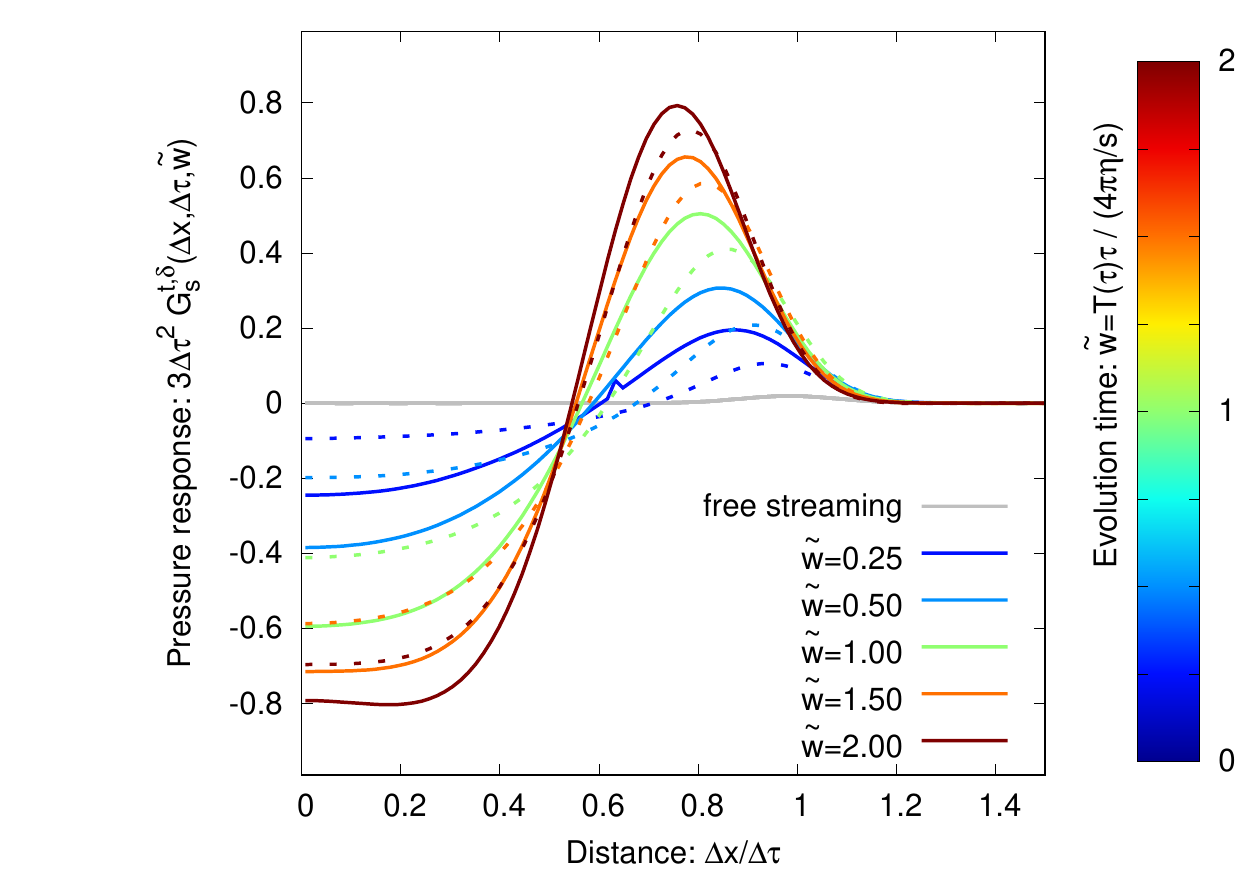}
\includegraphics[width=\textwidth]{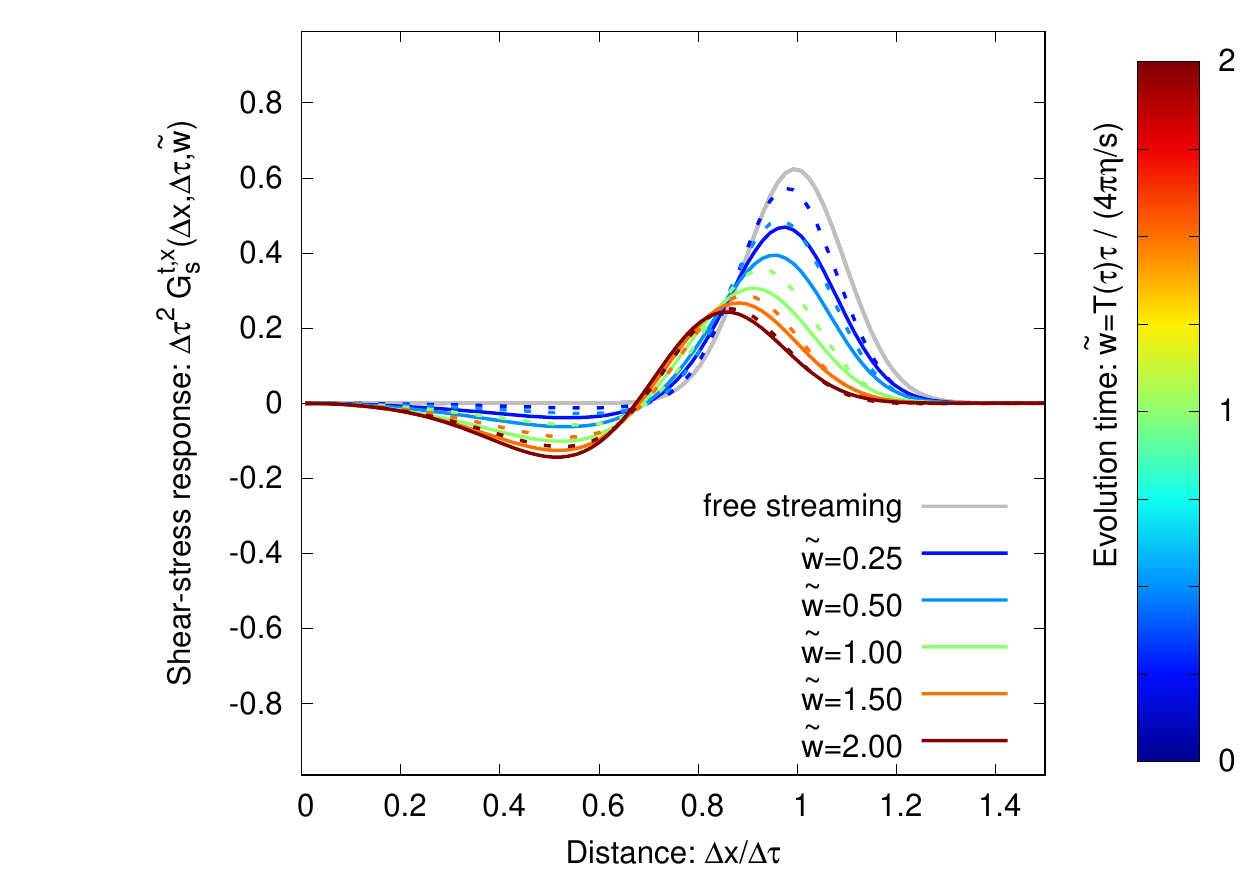}
\end{minipage}
\caption{\label{fig:EnergyResponseX} Evolution of the energy-momentum response to an initial energy perturbation in coordinate space, based on RTA (solid) and Yang-Mills kinetic theory (\kompost)(dotted) \cite{Kurkela:2018vqr} .  Different curves in each panel correspond to different evolution times $T(\tau)\tau/(4\pi \eta/s)$; different panels show the response of the different components of the energy-momentum tensor as a function of the propagation distance $|\xt-\xt_0|/(\tau-\tau_0)$.}
\end{center}
\end{figure*}

\begin{figure*}[t!]
\begin{center}
\begin{minipage}{0.49\textwidth}
\includegraphics[width=\textwidth]{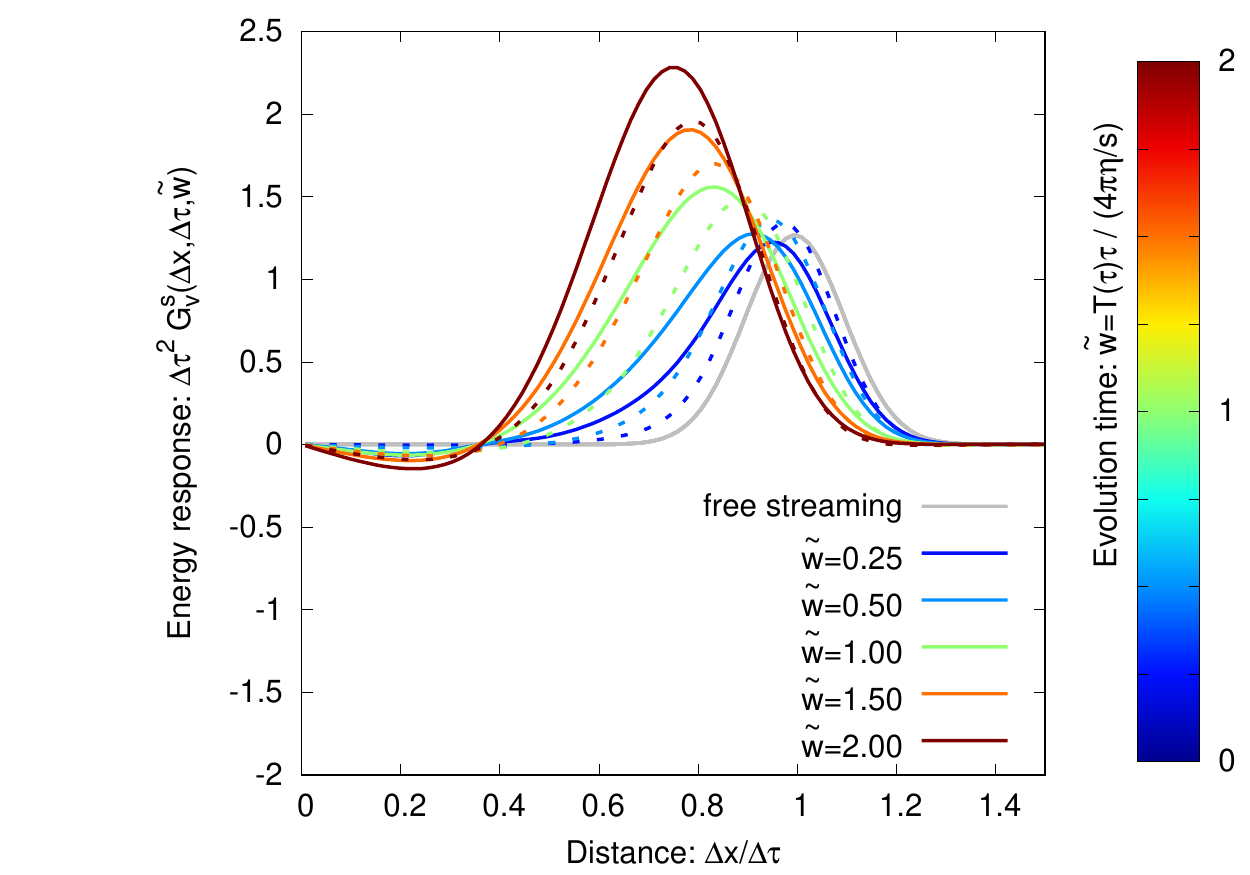}
\includegraphics[width=\textwidth]{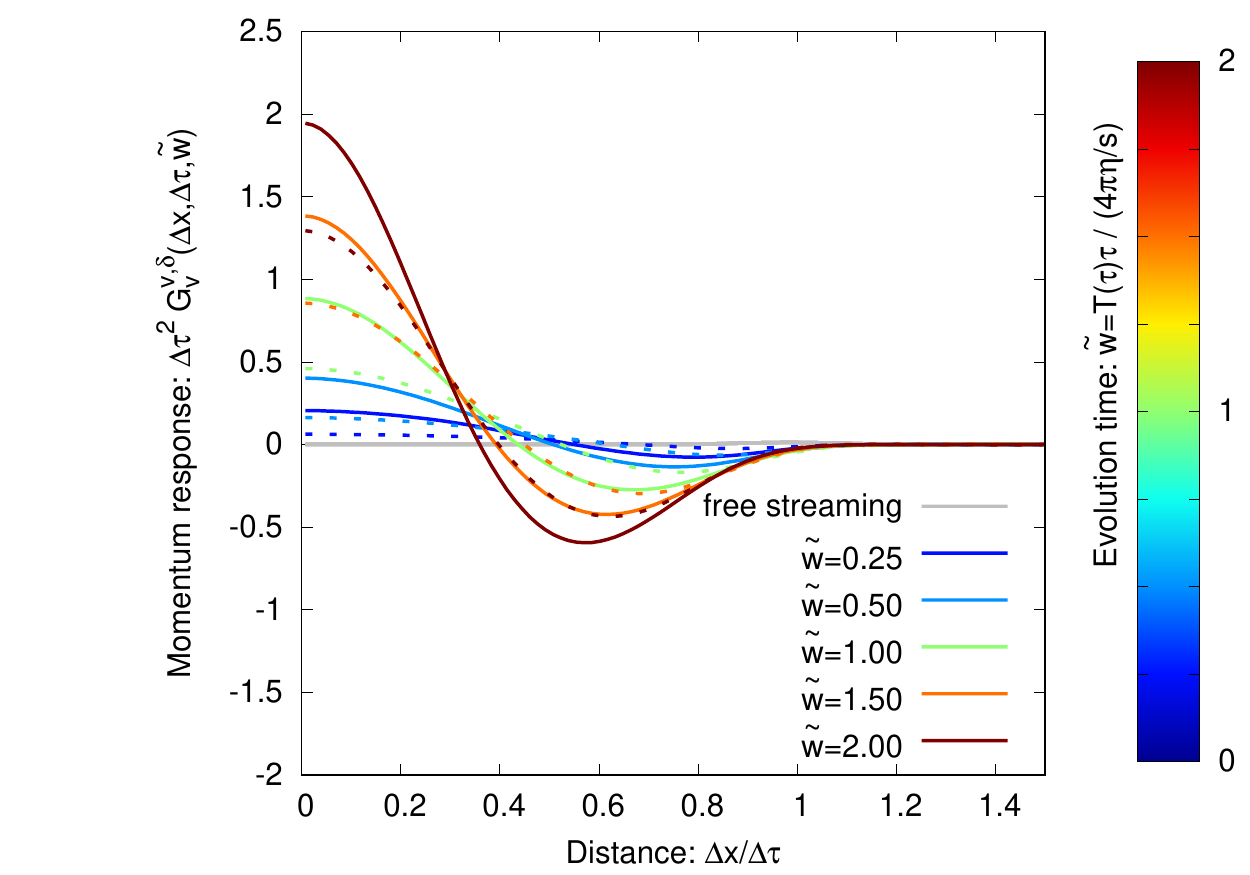}
\includegraphics[width=\textwidth]{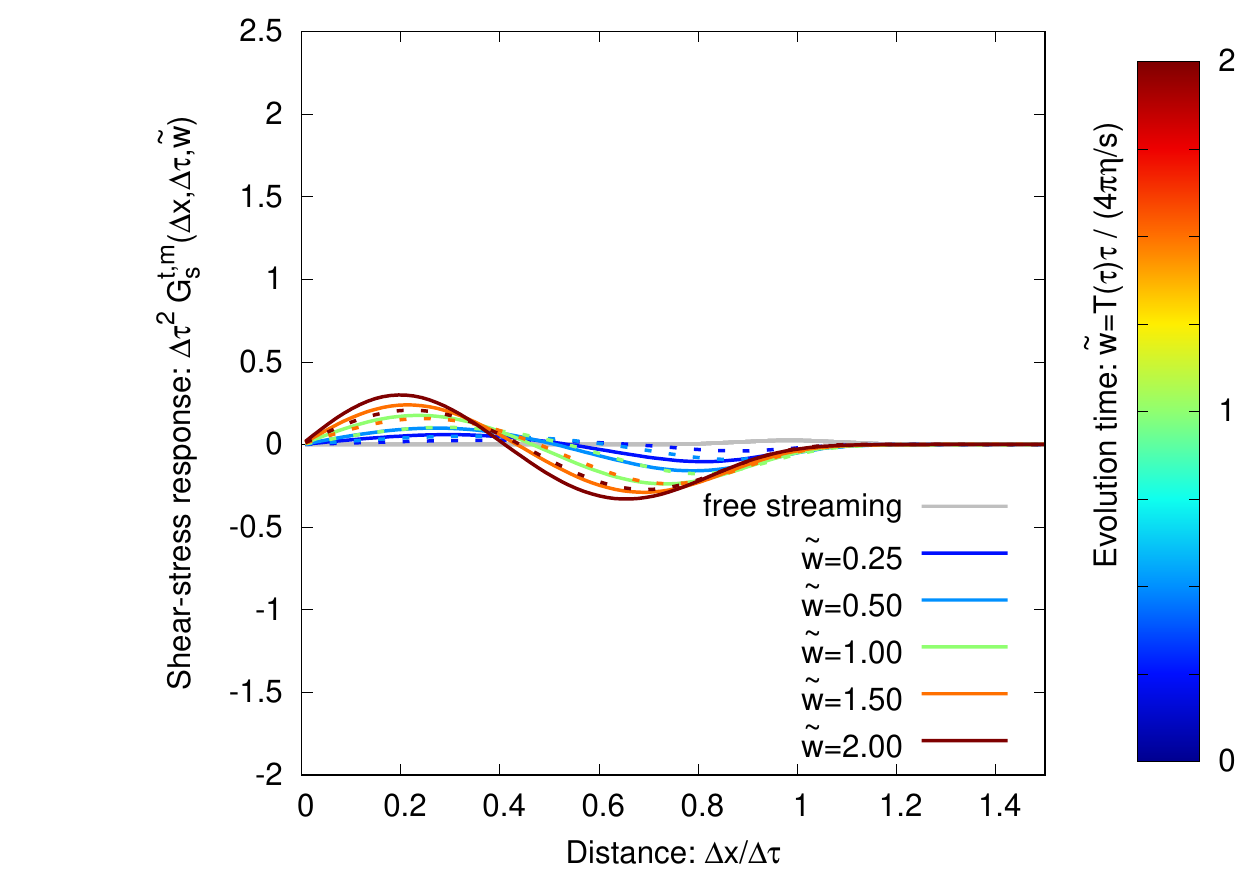}
\end{minipage}
\begin{minipage}{0.49\textwidth}
\includegraphics[width=\textwidth]{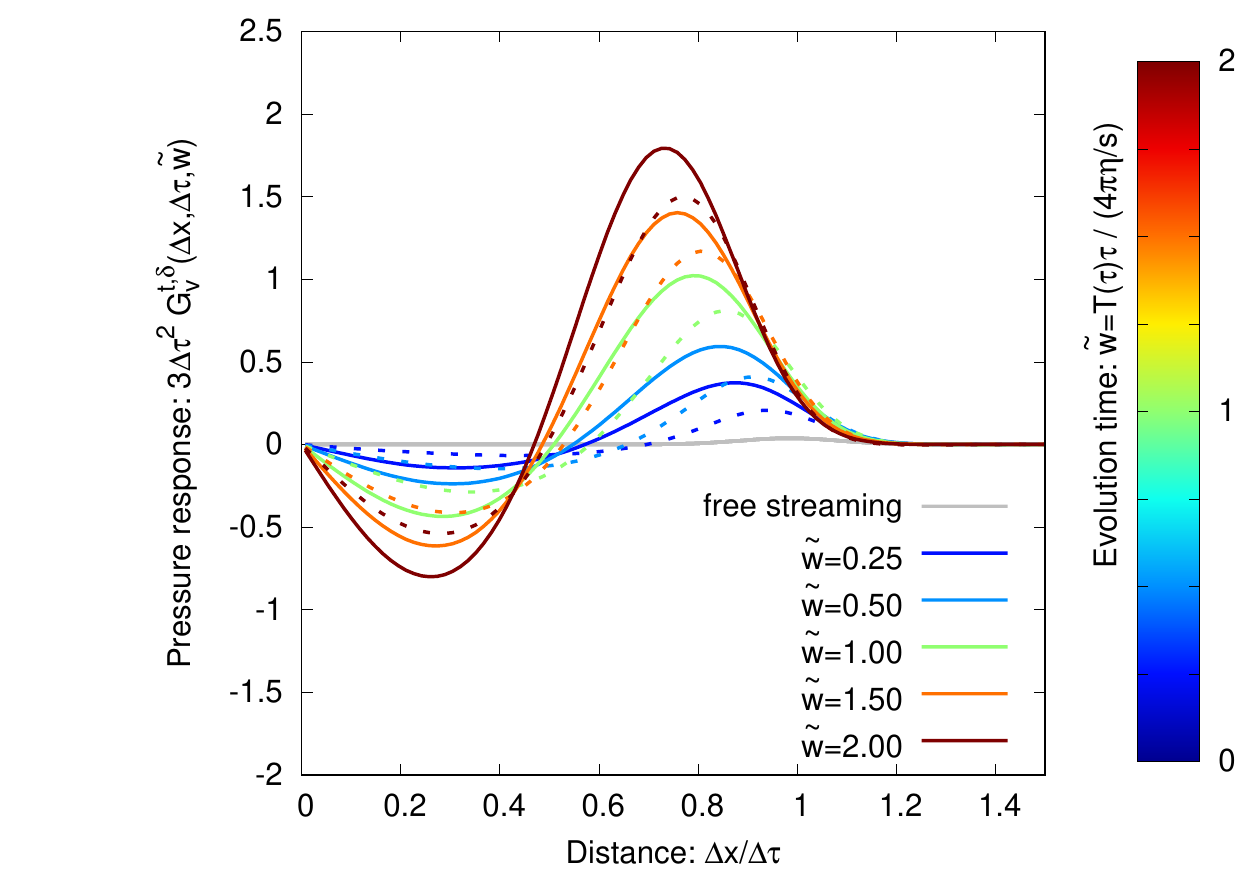}
\includegraphics[width=\textwidth]{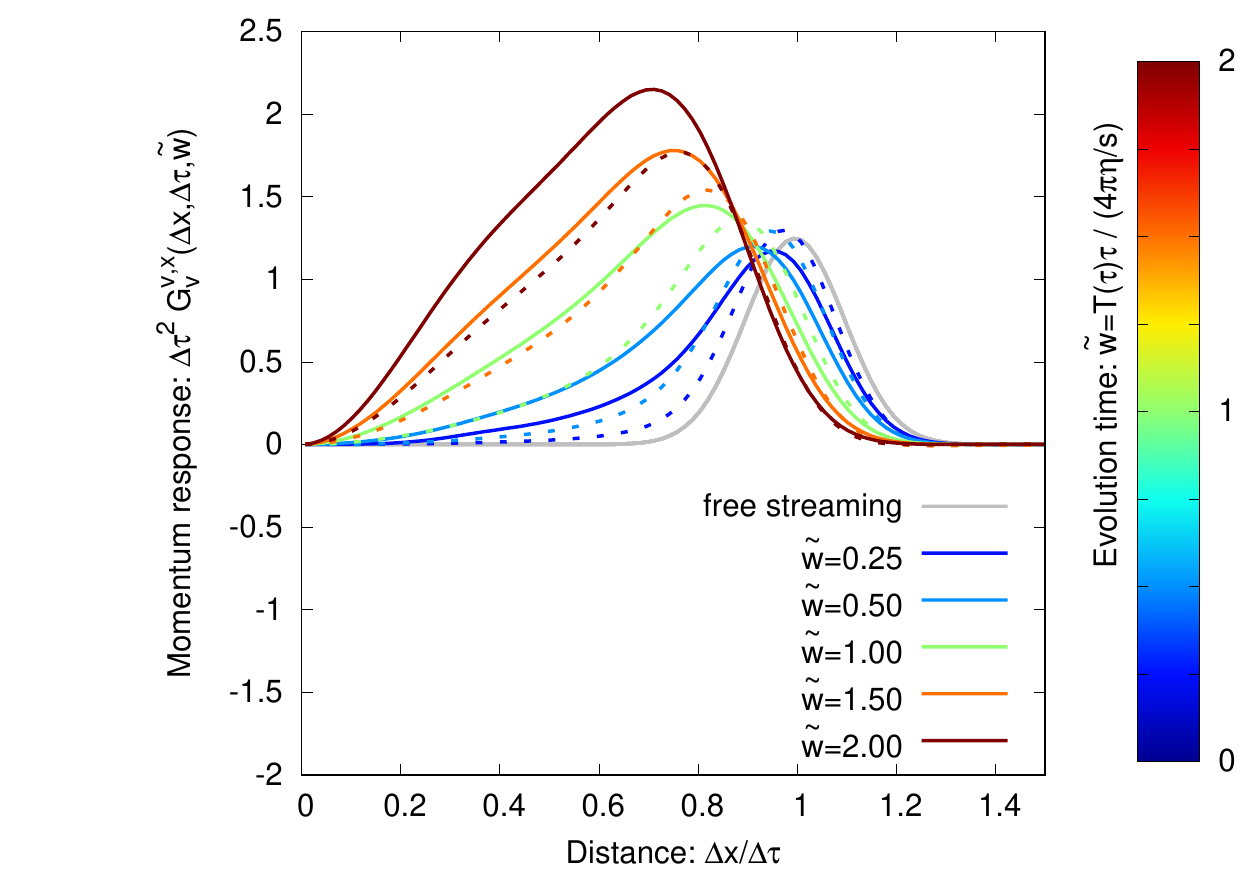}
\includegraphics[width=\textwidth]{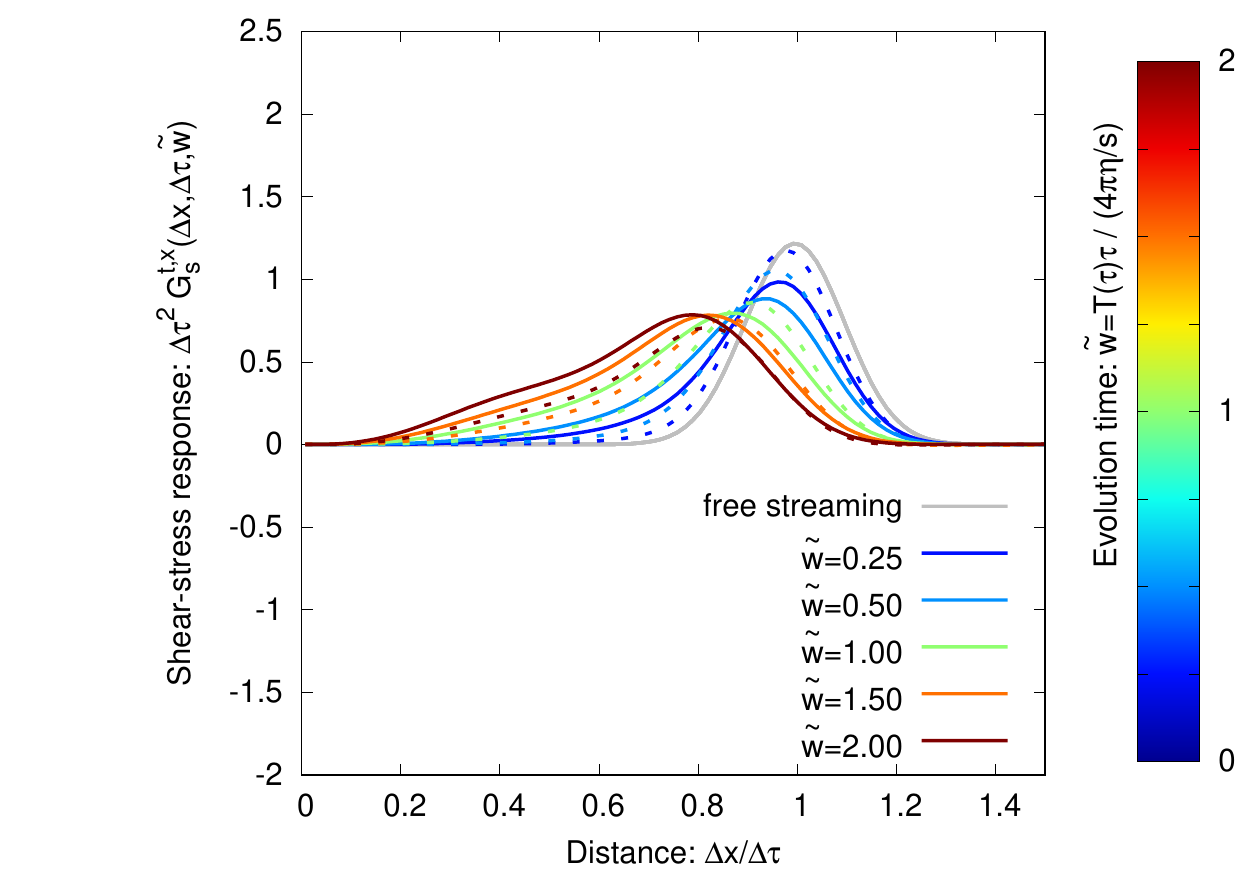}
\end{minipage}
\caption{\label{fig:MomentumResponseX} Evolution of the energy-momentum response to an initial momentum perturbation in coordinate space, based on RTA (solid) and Yang-Mills kinetic theory (\kompost)(dotted) \cite{Kurkela:2018vqr}.  Different curves in each panel correspond to different evolution times $T(\tau)\tau/(4\pi \eta/s)$; different panels show the response of the different components of the energy-momentum tensor as a function of the propagation distance $|\xt-\xt_0|/(\tau-\tau_0)$.}
\end{center}
\end{figure*}

\subsection{Numerical results}
We will now present numerical results for the various response functions, focusing on the case of a conformal relaxation time. We follow the same strategy as for the background and solve a truncated set of evolution equations, checking that including higher order moments does not significantly alter the results. Noteably, we find that a rather large set of moments is required to achieve apparent convergence and we will present results for $l_{\rm max} =64$ in the following.

Our results for the evolution of the response functions are compactly summarized in Figs.~\ref{fig:EnergyResponseK} and \ref{fig:MomentumResponseK} , where we present the spectrum of perturbations $G^{\mu\nu}_{\tau\tau}(\kt,\tau-\tau_0,\tau/\tau_R)$ and $G^{\mu\nu}_{\tau i}(\kt,\tau-\tau_0,\tau/\tau_R)$ as a function of $|\kt|(\tau-\tau_0)$ for different values of the evolution time $\tilde{w}=T(\tau)\tau/(4\pi \eta/s)$. Different panels in Figs.~\ref{fig:EnergyResponseK} and \ref{fig:MomentumResponseK} show the results for the different ($\mu\nu$) components, decomposed in the tensor basis described above. In order to facilitate the interpretation, we have also labeled the various response functions according to the components of the energy-momentum tensor that they affect.

Starting from the initial free-streaming behavior at early times $T(\tau)\tau/(4\pi \eta/s) \ll 1$ discussed in Sec.~\ref{sec:InitialConditons}, one observes how towards later times the viscous damping of large $|\kt|(\tau-\tau_0)$ modes sets in, such that by the time the system enters the hydrodynamic regime $(\tilde{w}=T(\tau)\tau/(4\pi \eta/s) \sim 1)$ only the long wave-length modes survive. Since at early times the system is highly anisotropic, the longitudinal pressure is effectively zero and transverse perturbations initially propagate with a phase-velocity of nearly the speed of light. Subsequently, as the system becomes more and more isotropic the phase-velocity decreases and eventually approaching the speed of sound, which in Figs.~\ref{fig:EnergyResponseK} and \ref{fig:MomentumResponseK} results in a shift of the peaks towards larger values of the propagation phase $|\kt|(\tau-\tau_0)$. Strikingly, the qualitative behavior observed from Figs.~\ref{fig:EnergyResponseK} and \ref{fig:MomentumResponseK} is very similar to the results obtained in Yang-Mills kinetic theory in \cite{Kurkela:2018vqr}, albeit we find that in the relaxation time approximation the viscous damping of short wave length modes becomes efficient on a somewhat shorter time scale.

Even though some of the features of the evolution can be understood quite naturally in wave-number $(\kt)$ space, in practice one is mostly interested in the Green's functions  $G^{\mu\nu}_{\tau\tau}(\xt-\xt_0,\tau-\tau_0,\tau/\tau_R)$ in position space, which directly describe the physical response of the energy momentum tensor $\delta T^{\mu\nu}(\tau,\xt)$  to a localized initial energy perturbation $\delta T^{\tau\tau}(\tau_0,\xt_0)$ according to
\begin{eqnarray}
\label{eq:GreensFunctionXDef}
\frac{\delta T^{\mu\nu}(\tau,\xt)}{e(\tau)}=\frac{1}{2}\int_{\xt_0}~G^{\mu\nu}_{\alpha\beta}(\xt-\xt_0,\tau-\tau_0) \frac{\delta T^{\alpha\beta}(\tau_0,\xt_0)}{e(\tau_0)}\;. \nonumber \\
\end{eqnarray}
In practice, the coordinate space response can be obtained in a straightforward way via a set of Bessel-Fourier transforms of the response functions, and we use the \kompost software~\cite{KoMPoSTGit} to perform this task. We note that when implementing our results into \kompost, one needs to take into account that the results in \cite{Kurkela:2018vqr} are presented in terms of the time variable $x_S^{\rm Id} = T_{\rm Id}(\tau)\tau/(4\pi \eta/s)$, where $T_{\rm Id}(\tau)$ is defined via the asymptotic temperature according to $T_{\rm Id}(\tau)=\tau^{-1/3} \lim_{\tau\to\infty} (T \tau^{1/3})$. However, from the point of view of the RTA Boltzmann equation it is more natural to study the evolution as a function of the scaling variable $\tilde{w}= T(\tau)\tau/(4\pi \eta/s)$, where $T(\tau)$ denotes the equilibrium temperature obtained from $e(\tau)$ via Landau matching. Of course, the two quantities are related by $x_S^{\rm Id}= \tilde{w} \left( \frac{(\tau^{4/3} e)_{\infty}}{ \tau^{4/3} e(\tau)} \right)^{1/4}$ and we have taken this difference into account in all explicit comparisons presented in this paper. In order to provide an apples to apples comparison between results obtained within the relaxation time approximation and Yang-Mills kinetic theory, the different response functions are smeared out with the same smearing kernel $K_{\sigma}= \exp(-\kt^2(\tau-\tau_0)^2/2\sigma^2)$ as in \cite{Kurkela:2018vqr}, which for Yang-Mills kinetic theory results is necessary in order to stabilize the numerical Bessel-Fourier transform. 

Our results for the coordinate space response functions are shown in Figs.~\ref{fig:EnergyResponseX} and \ref{fig:MomentumResponseX}, where we display the various response functions as a function of the propagation distance $|\xt-\xt_0|/(\tau-\tau_0)$. Solid curves in Figs.~\ref{fig:EnergyResponseX} and \ref{fig:MomentumResponseX} show our results obtained from the Boltzmann equation in relaxation time approximation, which are compared to the results obtained in Yang-Mills kinetic theory from \cite{Kurkela:2018vqr} shown as dashed curves. Some of the most important features that can be immediately observed from Figs.~\ref{fig:EnergyResponseX} and \ref{fig:MomentumResponseX}, include the viscous broadening of the peaks due to the damping of high wave-number modes, as well as the shift of the peaks towards smaller values of $|\xt-\xt_0|/(\tau-\tau_0)$ associated with the aforementioned change of the effective (transverse) speed of sound, due to the increase of the longitudinal pressure. One also observes a decrease of the shear-stress response compared to the pressure response (for both energy and momentum perturbations), such that by $\tilde{w}= T(\tau)\tau/(4\pi \eta/s) \sim 1$ when the background evolution starts to be captured by viscous hydrodynamics, also the dissipative corrections to the perturbations become sub-leading.

While the qualitative behavior of the various response functions is quite similar for RTA and Yang-Mills kinetic theory, one clearly observes that the RTA shows a faster departure from the early free streaming behavior, leading to differences $\sim 10\%$ on the relevant time scales $\tilde{w}=T(\tau)\tau/(4\pi \eta/s)\sim 1$ where viscous hydrodynamics becomes applicable. Even though these differences appear to be rather small, it would nevertheless be interesting to explore to what extent such differences in the early-time non-equilibrium dynamics can manifest themselves in final state observables in high-energy heavy-ion collisions.
\section{Conclusions \& Outlook}
\label{sec:concl}

We derived a new method to calculate non-equilibrium Green's function of the energy momentum tensor based on moment equations of kinetic equations for linearized perturbations. Due to the particularly simple structure of the relaxation time approximation considered in this work, we obtained a closed set of moment equations for the evolution of the dimension four moments $C_{l}^{m}$ and $\delta C_{l}^{m}$, which are relevant to study the evolution of the energy-momentum tensor, that is determined by the lowest order ($\ell \leq 2$) moments. Even though, for more complex interactions, a truncation at the level of dimension four operators is no longer sufficient to obtain a closed set of evolution equations, we naturally believe that by including higher order operators our method can be extended to systems with more complex interactions, thereby generalizing the usual moment method at level of perturbations.

Based on the evolution equations for the moments $C_{l}^{m}$ and $\delta C_{l}^{m}$ in Eqns.~(\ref{eq:BGClmEvo}) and (\ref{eq:dC_EOM}), we studied the evolution of average energy-momentum tensor in Bjorken flow, as well as the out-of-equilibrium linear response of the system to initial energy and momentum perturbations in the transverse plane. By truncating the infinite hierarchy of moment equations at large finite order, we obtained numerical solutions for the evolution of the non-equilibrium background and the Green's functions. When comparing our results to previous calculations of \kompost in Yang-Mills theory~\cite{Kurkela:2018vqr}, we found a striking similarity between the different theories. Even though the macroscopic differences between the two microscopic calculations are only at the ten percent level, it would be interesting to explore to what extent these can affect concrete observables, such as e.g. the flow harmonics $v_{n}$, the charged particle multiplicity $dN_{\rm ch}/d \eta$ or the transverse energy $dE_{\bot}/d \eta$. Since there are first hints that in particular the charged particle multiplicity $dN_{\rm ch}/d \eta$, may be a rather sensitive measure of the entropy production during the pre-equilibrium phase~\cite{Giacalone:2019ldn}, this remains an interesting question which we expect to be addressed in more detail in future studies.

Besides our numerical studies, we also found that various conformal scaling features, which have been empirically observed in \cite{Kurkela:2018vqr}, can be directly seen at level of the equations of motion for the moments  $\delta C_{l}^{m}$, and we expect that further analytic insights into the structure of the non-equilibrium Green's functions. Specifically, it would be interesting to further explore for example the early and late time asymptotics of the Green's functions based on this formulation. Furthermore, one can also investigate to what extent the highly non-trivial evolution of the Green's functions can be captured by "renormalized transport coefficients" as suggested in various works \cite{Blaizot:2017ucy,Behtash:2018moe,Denicol:2020eij} studying the highly symmetric Bjorken flow.

Beyond such analytic insights, it would also interesting to further systematically extent the pre-equilibrium description in terms of non-equilibrium Green's functions. Since the numerical solution of the moment equations is comparatively straightforward, the Boltzmann equation in relaxation time approximation provides an ideal testing ground for such ideas. Specifically, with the methodology developed in this paper it should be comparatively straightforward to extent the description to include also longitudinal fluctuations, investigate Green's functions for higher order moments or study the effect of additional conserved charges, all of which are rather challenging tasks within a QCD kinetic  description.  By explicitly comparing the linearized description in terms of non-equilibrium Green's functions to full numerical solutions of the RTA Boltzmann equation, one could also obtain additional insights into the reliability and breakdown of the linearized description and potentially improve the range of applicablity to describe small collision systems.\\


\acknowledgments
We thank A. Behtash for his involvement in this project during its early stages. MM and SK were supported in part by the US Department of Energy grant DE-FG02-03ER41260. MM is also supported by the BEST (Beam Energy Scan Theory) DOE Topical Collaboration. SS, PP and SO are supported by the Deutsche Forschungsgemeinschaft (DFG, German Research Foundation) through the  CRC-TR 211 ``Strong-interaction matter under extreme conditions" under Project number 315477589.

\appendix
\section{Identities for spherical harmonics \& associated Legendre polynomials}
\label{sec:SphericalHarmonicIdentities}
Below we summarize some of the identities used to derive the evolution equations for the background moments $C^{m}_{l}$ and the linearized perturbations $\delta C_{l}^{m}$. Specifically, for the evolution equations of the background moments we make use of the identities
\begin{eqnarray}
(1-x^2) \frac{d}{dx} P_{l}^{m}(x) = \Delta_{l,-}^{m} P_{l-1}^{m}(x) + \Delta_{l,+}^{m} P_{l+1}^{m}\;, \qquad  
\end{eqnarray}
and
\begin{eqnarray}
x P_{l}^{m}(x)= \xi_{l,-}^{m} P_{l-1}^{m}(x) + \xi_{l,+}^{m} P_{l+1}^{m}(x)\;,
\end{eqnarray}
as well as
\begin{eqnarray}
x^2 P_{l}^{m}(x)=\xi_{l,-2}^{(2),m} P_{l-2}^{m}(x) + \xi_{l,0}^{(2),m}  P_{l}^{m}(x)  +   \xi_{l,+2}^{(2),m} P_{l+2}^{m}(x)\;, \nonumber \\
\end{eqnarray}
to derive the relation
\begin{eqnarray}
&&\left[\left(\frac{1}{3}-x^2\right) - x (1-x^2) \frac{d}{dx} \right]  P_{l}^{m}(x)= \\
&& \qquad a_{l,-2}^{m}  P_{l-2}^{m}(x) + a_{l,0}^{m}  P_{l}^{m}(x) + a_{l,+2}^{m}  P_{l+2}^{m}(x)\;. \nonumber
\end{eqnarray}
such that
\begin{eqnarray}
\tau \partial_{\tau} C_{l}^{m} &=& b_{l,-2}^{m}  C_{l-2}^{m} + b_{l,0}^{m}  C_{l}^{m} + b_{l,+2}^{m}  C_{l+2}^{m}  \\
&&+  \int \frac{d^2p}{(2\pi)^2} \int \frac{dp_{\varsigma}}{(2\pi)} \left( \tau^{1/3} p^{\tau} \right) Y_{l}^{m}(\phi,\theta) \tau \partial_{\tau}  f(p)\;. \nonumber
\end{eqnarray}
which is the identity used in the main text. Similarly, making use of the relations
\begin{eqnarray}
P_{l}^{-m}(x)=\sigma_{l}^{m} P_{l}^{m}(x)\;, \qquad \sigma_{l}^{m}=(-1)^{m} \frac{(l-m)!}{(l+m)!}\;, \nonumber \\
\end{eqnarray}
along with
\begin{eqnarray}
\sqrt{1-x^2} P_{l}^{m}=\frac{1}{2l+1} \Big(P_{l-1}^{m+1}(x) - P_{l+1}^{m+1}(x) \Big)\;, \nonumber \\
\end{eqnarray}
we can evaluate the additional terms
\begin{eqnarray}
\sin(\theta) e^{+i\phi} Y_{l}^{m}(\phi,\theta) &=& u_{l,-}^{m}Y_{l-1}^{m+1}(\phi,\theta) + u_{l,+}^{m}Y_{l+1}^{m+1}(\phi,\theta) \;, \nonumber \\
\\
\sin(\theta) e^{-i\phi} Y_{l}^{m}(\phi,\theta) &=&  d_{l,-}^{m}Y_{l-1}^{m-1}(\phi,\theta) + d_{l,+}^{m}Y_{l+1}^{m-1}(\phi,\theta) \;, \nonumber \\
\end{eqnarray}
which arise in the evolution equations for the linearized energy-momentum perturbations.  Below we list the coefficients entering the above identities
\begin{eqnarray}
\Delta_{l,-}^{m} &=&\frac{(l+1)(l+m)}{2l+1}\;, \\
\Delta_{l,+}^{m} &=& -\frac{l (l-m+1)}{2l+1}\;, \nonumber
\end{eqnarray}
\begin{eqnarray}
\xi_{l,-}^{m}&=&\frac{l+m}{2l+1}\;, \\
\xi_{l,+}^{m}&=&\frac{l-m+1}{2l+1}\;, \nonumber
\end{eqnarray}
\begin{eqnarray}
\xi_{l,-2}^{(2),m}&=&\xi_{l,-}^{m} \xi_{l-1,-}^{m}\;, \\
\xi_{l,0}^{(2),m}&=& \left(\xi_{l,-}^{m} \xi_{l-1,+}^{m} + \xi_{l,+}^{m} \xi_{l+1,-}^{m}\right)\;, \nonumber \\
\xi_{l,+2}^{(2),m}&=&\xi_{l,+}^{m} \xi_{l+1,+}^{m} \;, \nonumber
\end{eqnarray}
\begin{eqnarray}
a_{l,-2}^{m}&=& -\xi_{l,-2}^{(2),m} - \Delta_{l,-}^{m} \xi_{l-1,-}^{m}\;, \\
a_{l,0}^{m}&=&\frac{1}{3} - \xi_{l,0}^{(2),m} -  \Delta_{l,-}^{m} \xi_{l-1,+}^{m} -  \Delta_{l,+}^{m} \xi_{l+1,-}^{m}\;,  \nonumber \\
a_{l,+2}^{m}&=& -\xi_{l,+2}^{(2),m} - \Delta_{l,+}^{m} \xi_{l+1,+}^{m}\;. \nonumber
\end{eqnarray}
\begin{eqnarray}
b_{l,-2}^{m}&=&a_{l,-2}^{m}  \frac{y_{l}^{m}}{y_{l-2}^{m}}\;, \\
b_{l,0}^{m}&=&a_{l,0}^{m}\;,  \nonumber \\
b_{l,+2}^{m}&=& a_{l,+2}^{m} \frac{y_{l}^{m}}{y_{l+2}^{m}}\;. \nonumber
\end{eqnarray}
\begin{eqnarray}
u_{l,-}^{m}&=&+\frac{y_{l}^{m}}{(2l+1)y_{l-1,m+1}}\;, \\
u_{l,+}^{m}&=&-\frac{y_{l}^{m}}{(2l+1)y_{l+1,m+1}}\;,  \nonumber 
\end{eqnarray}
\begin{eqnarray}
d_{l,-}^{m}&=&+\frac{y_{l}^{m}}{(2l+1)y_{l-1,m-1} \sigma_{l}^{m}\sigma_{l-1}^{-m+1}}\;, \\
d_{l,}^{m}&=&-\frac{y_{l}^{m}}{(2l+1)y_{l+1,m-1}\sigma_{l}^{m}\sigma_{l+1}^{-m+1}}\;, \nonumber
\end{eqnarray}
which upon further simplifications yield the results quoted in the main text.


\bibliography{non_equilibrium_green}
\end{document}